\documentclass[12pt,a4paper]{article}

\usepackage{amssymb}
\usepackage{amsmath}
\usepackage{setspace}
\usepackage{indentfirst}

\usepackage{graphicx}

\usepackage{url}

\begin{document}

\begin{titlepage}
\begin{center}

\vspace*{25mm}

\begin{spacing}{1.7}
{\LARGE\bf Reionization process dependence of the ratio of CMB polarization power spectra at low-$\ell$}
\end{spacing}

\vspace*{25mm}

{\large
Noriaki Kitazawa
}
\vspace{10mm}

Department of Physics, Tokyo Metropolitan University,\\
Hachioji, Tokyo 192-0397, Japan\\
e-mail: noriaki.kitazawa@tmu.ac.jp

\vspace*{25mm}

\begin{abstract}
We investigate
 how much the ratio of cosmic microwave background (CMB) polarization power spectra $C^{BB}_\ell/C^{EE}_\ell$
 at low-$\ell$ ($\ell \lesssim 10$) depends on the process of reionization.
Both such low-$\ell$ B-mode and E-mode polarization powers
 are dominantly produced by Thomson scattering of CMB photons off the free electrons
 which are produced in the process of reionization.
Since the reionization should be finished until at least the redshift $z \simeq 6$
 and the low-$\ell$ polarization powers are produced at late time,
 the ratio is rather insensitive by the ionization process at higher redshifts,
 but it is sensitive to the value of optical depth.
The value of the ratio at $\ell=2$, however,
 is almost insensitive to the reionization process including the value of optical depth,
 and the value is approximately half of the value of tensor-to-scalar ratio.
This fact can be utilized for future determination of tensor-to-scalar ratio
 in spite of the ambiguity due to cosmic variance.
\end{abstract}

\end{center}
\end{titlepage}

\doublespacing

\section{Introduction}
\label{sec:introduction}

The signal of low-$\ell$ ($\ell \lesssim 10$) B-mode polarization
 by the future probe in space, LiteBIRD \cite{Hazumi:2021yqq} for example,
 is expected to be a promising evidence of the primordial tensor perturbations.
The higher-$\ell$ B-mode signal by the present and future probes on the ground also can be such an evidence,
 but the dominant contribution from the conversion of E-mode polarization
 by the gravitational lensing effect should be subtracted.
Although there is no such a contamination for low-$\ell$ B-mode,
 the magnitude of the power of low-$\ell$ B-mode polarization strongly depends on the reionization process,
 since it is mainly produced by Thomson scattering of CMB photons off the free electrons
 which are produced in the process of reionization.

Establishing precise knowledge of the reionization process is underway. 
Much effort has been devoted to determine the form of the ionization function, $x_e(z)$,
 which describes the rate of ionization of neutral hydrogen (and helium)
 at the time corresponding to the redshift $z$.
The ionization is considered to be happened by the ultraviolet light from stars and other objects,
 but the standard precise understanding of the star and galaxy formation has not been established yet.
Therefore,
 there has been mainly two ways to determine the form of ionization function from cosmological observations.
One way is that
 we assume some physically motivated model of the ionization function which includes some parameters,
 and fit these parameters with Planck CMB data
 (see \cite{Qin:2020xrg,Ahn:2020btj,Wu:2021kch} for recent efforts).
Another way is the model independent analysis in which
 we introduce some general representation for the ionization function which also includes some parameters,
 and fit these parameters with Planck CMB data
 (see \cite{Heinrich:2016ojb,Miranda:2016trf,Paoletti:2020ndu,Heinrich:2021ufa} for recent efforts).
Beyond Planck CMB data,
 recent observations of Lyman-$\alpha$ absorption lines in the spectra from distant quasars
 indicate that the reionization process should be ended (95\% ionization) at least 
 $z \simeq 5.2 \sim 5.6$ \cite{Choudhury:2020vzu,Qin:2021gkn}.
The analyses with both Planck CMB data and astrophysical observations
 are given in \cite{Hazra:2019wdn,Paoletti:2021gzr}.

However it is difficult to obtain a standard precise form of the ionization function,
 since it requires more information from cosmological observations.
For example, Pop-III stars may give a long tail in the ionization function beyond $z>15$,
 which is supported in \cite{Ahn:2020btj,Heinrich:2016ojb},
 but not supported in \cite{Paoletti:2020ndu,Planck:2016mks} on the other hand.
In \cite{Wu:2021kch}
 it is even claimed that such a long tail can not be constrained by Planck CMB data at low-$\ell$. 
Therefore,
 the main results of the previous analyses on the reionization process
 have been focused on the value of optical depth, $\tau$,
 which is proportional to the integration of the ionization function.
In the latest analysis by Planck collaboration \cite{Planck:2018vyg}
 ``instantaneous reionization'',
 in which the ionization function is essentially described by $\tanh$ function,
 is assumed to obtain the value of optical depth.
Since the values of optical depth by many previous analyses
 are consistent with the value given in this latest analysis by Planck collaboration,
 we simply use this model of instantaneous reionization
 as a test model for numerical calculations in this article.

The E-mode polarization powers at higher-$\ell$
 have already measured by the probes in space and on the ground.
The higher-$\ell$ E-mode polarizations are produced in the period of recombination,
 and the magnitude of power quickly decreases for smaller $\ell$.
The low-$\ell$ E-mode polarization is dominantly produced also
 by Thomson scattering of CMB photons off the free electrons
 which are produced in the process of reionization.
Therefore,
 it is naively expected that the ratio of powers, $C^{BB}_\ell/C^{EE}_\ell$, at low-$\ell$
 is not affected so much by the detail of the reionization process.
Especially for very small $\ell$
 it is expected that the ratio is almost independent from the detailed process of reionization,
 since the reionization should be finished until at least $z \simeq 6$
 and such polarizations are produced at the late time.  
We will see that this observation is true especially in case of $\ell=2$.
Other ratios like $C^{BB}_\ell/C^{TT}_\ell$ and $C^{BB}_\ell/C^{TE}_\ell$ do not have this property,
 since the origin of temperature perturbations is different from that of polarizations at small $\ell$.

In the next section
 we analytically investigate the reionization dependence of this ratio
 in the approximation of large wave length limit of scalar and tensor perturbations
 \cite{Polnarev:1985,Harari:1993nb,Zaldarriaga:1995gi,Ng:1993fx,Ng:1994sv,Kamionkowski:1996ks,Keating:1997cv,
 Zhang:2005nv,Cabella:2004mk,Pritchard:2004qp,Kitazawa:2019fzc,Kitazawa:2020qdx}.
The value of the ratio at $\ell=2$ is estimated to be the half of tensor-to-scalar ratio.
In section \ref{sec:numerical}
 we numerically calculate the ratio using CAMB cade \cite{Lewis:1999bs}
 changing the form of the ionization function and the value of optical depth.
The value of the ratio depends on the reionization process,
 especially on the value of optical depth, in the level of 10\% for $\ell=10$.
However the dependence becomes smaller for smaller $\ell$,
 and the value at $\ell=2$ is almost independent from the reionization process.
We find that the value at $\ell=2$
 is actually equal to the half of tensor-to-scalar ratio in good precision.
In the last section we conclude with some discussion.

\section{Analytic arguments}
\label{sec:analytic}

The polarization powers of CMB perturbations at low-$\ell$ are obtained by solving the Boltzmann equations
 which describe CMB photons propagating in scalar or tensor perturbations of the background metric
 with Thomson scattering off the free electrons which are produced in the process of reionization.
These Boltzmann equations have been introduced and analytically developed in
 \cite{Polnarev:1985,Harari:1993nb,Zaldarriaga:1995gi,Ng:1993fx,Ng:1994sv,Kamionkowski:1996ks,Keating:1997cv,
 Zhang:2005nv,Cabella:2004mk,Pritchard:2004qp,Kitazawa:2019fzc,Kitazawa:2020qdx},
 and in this article we follow the notation and results in \cite{Kitazawa:2019fzc,Kitazawa:2020qdx}.
The effect of the scattering changing in conformal time $\eta$ is described by a function
\begin{equation}
 g(\eta) = \sigma_T n_e(\eta) a(\eta),
\end{equation}
 where $\sigma_T$ is the cross section of Thomson scattering,
 $n_e(\eta)$ is the free electron density and $a(\eta)$ is the scale factor.
The ionization function is defined as $x_e(\eta) \equiv n_e(\eta) a(\eta)^3 / n_{{\rm H}0}$,
 where $n_{{\rm H}0}$ is the neutral hydrogen density at present.
The normalization of the scale factor is taken as $a(\eta_0)=1$ assuming flat space-time,
 and the relation between the conformal time $\eta$ and the redshift $z$ is given as $1+z=1/a(\eta)$.
The optical depth is given by
\begin{equation}
 \tau = \int_0^\infty d\eta \, g(\eta)
      = \sigma_T \, n_{{\rm H}0} \int_0^\infty dz \frac{(1+z)^2}{H(z)} x_e(z),
\end{equation}
 where $H(z)$ is the Hubble parameter as a function of redshift $z$.
In the long wavelength limit (or the tight coupling limit) $k \gg g$,
 where $k$ is the wave number of the background perturbations,
 we can obtain approximate analytic formulae for angular power spectra of polarizations.
Since the long wavelengths correspond to small $\ell$'s in the angular power spectra,
 the formulae are applicable for small $\ell \lesssim 10$.

The analytic formulae for polarization power spectra
 in long wavelength limit (or small $\ell$ limit) are obtained as follows
 (see \cite{Kitazawa:2019fzc,Kitazawa:2020qdx} for full process to obtain the following formulae).

The E-mode power spectrum is obtained as follows.
\begin{equation}
 C_\ell^{\rm EE}
  = \int \frac{d^3k}{(2\pi)^3} \, \frac{1}{2\ell+1}
    \sum_{m=-\ell}^\ell \vert a^{\rm E}_{k,\ell m} \vert^2
\label{power-E}
\end{equation}
 with
\begin{equation}
\begin{split}
 a^{\rm E}_{k,\ell m} =
   &- \delta_{m0} T_0 \sqrt{\pi(2\ell+1)} \sqrt{\frac{2\ell(\ell-1)}{(\ell+1)(\ell+2)}} \beta^{\rm E}_{k,\ell}
\\
   &+ \delta_{m0} T_0 \sqrt{\pi(2\ell+1)} \sqrt{\frac{2\ell(\ell-1)}{(\ell+1)(\ell+2)}}
      \sum_{n=1}^{[\ell/2]} \frac{2(\ell-2n)-1}{\ell(\ell-1)/2} \beta^{\rm E}_{k,\ell-2n},
\end{split}
\label{amp-spherical-E}
\end{equation}
 where
\begin{equation}
\begin{split}
 \beta^{\rm E}_{k,\ell} \simeq
 \int_0^{\eta_0} d\eta' \, g(\eta') \, G^S_k(\eta') \, i^\ell
 \Bigg\{
  &\frac{\ell(\ell-1)}{(2\ell+1)(2\ell-1)} \, j_{\ell-2}(k(\eta'-\eta_0))
\\
 &+ \frac{2(\ell^2+\ell-1)}{(2\ell+3)(2\ell-1)} \, j_\ell(k(\eta'-\eta_0))
\\
 &+ \frac{(\ell+2)(\ell+1)}{(2\ell+3)(2\ell+1)} \, j_{\ell+2}(k(\eta'-\eta_0))
 \Bigg\}
\end{split}
\label{beta-E}
\end{equation}
 describes the magnitude of E-mode polarization component in the Boltzmann equation. 
Here, $j_\ell(x)$ is the spherical Bessel function
 and $G^S_k$ is so called source function for E-mode which is obtained in long wavelength limit as
\begin{equation}
 G^S_k(\eta) \simeq \frac{{\cal R}^0_k}{300} \, k^2(\eta^2-\eta_{\rm ion}^2)
\label{source-E0}
\end{equation}
 for long wavelength $k\eta < 1$,
 where $\eta_{\rm ion}$ is the conformal time for reionization to start,
 namely, $g(\eta < \eta_{\rm ion})$ is negligible, and  
\begin{equation}
 P_{\cal R}(k) = \frac{k^3}{2\pi^2} \vert {\cal R}^0_k \vert^2
  = A_S \left( \frac{k}{k_{\rm pivot}} \right)^{n_s-1}
\end{equation}
 is the primordial scalar perturbation
 with spectral index $n_s$
 and the magnitude $A_S$ at a pivot scale $k_{\rm pivot}$.
Therefore, in good approximation with $n_s \simeq 1$
\begin{equation}
 G^S_k(\eta) \simeq \frac{1}{300} \sqrt{\frac{2\pi^2 A_S}{k^3}} \, k^2(\eta^2-\eta_{\rm ion}^2).
\label{source-E}
\end{equation}
Note that
 the time evolution equation for scalar perturbations has to be solved to obtain eq.~(\ref{source-E0}).

The B-mode power spectrum is obtained as follows.
\begin{equation}
 C_\ell^{\rm BB}
  = 2 \int \frac{d^3k}{(2\pi)^3} \, \frac{1}{2\ell+1}
      \sum_{m=-\ell}^\ell \vert a^{\rm B}_{k,\ell m} \vert^2
\label{power-B}
\end{equation}
 with
\begin{equation}
 a^{\rm B}_{k,\ell m} =
  i \frac{T_0}{4} \left( \delta_{m \, 2} - \delta_{m \, -2} \right)
    \sqrt{\frac{2\pi}{2\ell+1}}
    \Big\{
     (\ell+2) \beta^{\rm B}_{k,\ell-1} + (\ell-1) \beta^{\rm B}_{k,\ell+1}
    \Big\},
\label{amp-spherical-B}
\end{equation}
 where
\begin{equation}
 \beta^{\rm B}_{k,\ell} \simeq
 \int_0^{\eta_0} d\eta' \, g(\eta') \, G^T_k(\eta') \, i^\ell \, j_\ell(k(\eta'-\eta_0))
\label{beta-B}
\end{equation}
 describes the magnitude of B-mode polarization component in the Boltzmann equation. 
Here, $G^T_k$ is so called source function for B-mode which is obtained in long wavelength limit as
\begin{equation}
 G^T_k(\eta) \simeq - \frac{1}{10} \left( {\cal D}_k(\eta) - {\cal D}_k(\eta_{\rm ion}) \right)
\end{equation}
 with
\begin{equation}
 {\cal D}_k(\eta) =
  \sqrt{\frac{2\pi^2 A_T}{k^3} \left( \frac{k}{k_{\rm pivot}} \right)^{n_t}}
  \times 3 \sqrt{\frac{\pi}{2}} (k\eta)^{-2/3} J_{3/2}(k\eta),
\end{equation}
 which is the solution of the time evolution equation of tensor perturbations,
 where $J_\nu(x)$ is the Bessel function and
\begin{equation}
 {\cal P}_h(k) = \frac{1}{4} {\cal P}_T(k) = A_T \left( \frac{k}{k_{\rm pivot}} \right)^{n_t}
\end{equation}
 is the primordial tensor perturbation
 with spectral index $n_t$ and the magnitude $A_T$ at a pivot scale $k_{\rm pivot}$.
The tensor-to-scalar ratio of the primordial perturbations is defined as
\begin{equation}
 r \equiv \frac{4A_T}{A_S}.
\label{tensor-to-scalar}
\end{equation}
For long wavelength $k\eta < 1$ with $n_t=0$
\begin{equation}
 G^T_k(\eta) \simeq \frac{1}{300} \sqrt{\frac{2\pi^2 A_T}{k^3}} \, k^2(\eta^2-\eta_{\rm ion}^2).
\label{source-B}
\end{equation}
Note that
 this equation is exactly the same of eq.~(\ref{source-E}) with replacing $A_S$ by $A_T$.

The physical meaning of the integrals in eqs.~(\ref{beta-E}) and (\ref{beta-B}) is the following.
Since the spherical Bessel function $j_\ell(x)$ is the dumping oscillating function as
\begin{equation}
 j_0(x) = \frac{\sin(x)}{x},
\quad
 j_1(x) = \frac{\sin(x)}{x^2} - \frac{\cos(x)}{x},
\end{equation}
 for example,
 it works roughly as a filter in the integral over $x$ to choose the region of $x \simeq \ell+1$.
In the integral of eq.~(\ref{beta-B})
 the spherical Bessel function works as a filter to choose a region of $k(\eta_0-\eta') \simeq \ell+1$,
 and the value $g(\eta')G^T_k(\eta')$ with $\eta' \simeq \eta_0 - (\ell+1)/k$ dominates the integral.
Note that
 the range of value of $\eta'$ is $\eta_{\rm ion} < \eta' < \eta_0$,
 since $g(\eta)$ is negligible for $\eta < \eta_{\rm ion}$. 
Because of the fact that $\eta_{\rm ion}$ is numerically the same order of magnitude of $\eta_0$,
 the value of wave number $k \simeq (\ell+1)/\eta_0$ is chosen in the integral,
 which indicates that smaller $\ell$ correspond to smaller $k$
 or long wavelength of tensor perturbations.
The integral in eq.~(\ref{beta-E}) can be understood in the same way, though there are three terms.
Note that for small $\ell$ the first term dominates,
 since the value of $g(\eta')G^S_k(\eta')$ is larger for larger $\eta'$.
Therefore,
 the dependence of $\beta^{\rm E}_{k,\ell}$ and $\beta^{\rm B}_{k,\ell}$
 on the reionization precess, which is specified by the shape of $g(\eta')$,
 is expected to be almost the same for small $\ell$
 with the same form of the functions $G^S_k(\eta')$ and $G^T_k(\eta')$.

The amplitudes of spherical harmonic expansions of polarization tensor,
 eqs.~(\ref{amp-spherical-E}) and (\ref{amp-spherical-B}),
 are described by $\beta^{\rm E}_{k,\ell}$ and $\beta^{\rm B}_{k,\ell}$, respectively.
In eq.~(\ref{amp-spherical-E}) the number of $\beta^{\rm E}_k$ increases with $\ell$,
 on the other hand in eq.~(\ref{amp-spherical-B}) the number of $\beta^{\rm B}_k$ is always two.
Therefore the values of these amplitudes
 should depend on the reionization process very differently for large $\ell$.
On the other hand for small $\ell$ the dependence is expected to be the same,
 because the form of formulae eqs.~(\ref{amp-spherical-E}) and (\ref{amp-spherical-B}) becomes similar
 with $\beta^{\rm E}_k$ and $\beta^{\rm B}_k$
 which have almost the same dependence on the reionization process.
Then,
 the dependence of the power spectra of eqs.~(\ref{power-E}) and (\ref{power-B})
 on the reionization process is expected to be the same for small $\ell$.

To be explicit, consider the the cases of $\ell=2$.
We estimate $C_2^{\rm EE}$ first.
The corresponding amplitude is
\begin{equation}
 a^{\rm E}_{k,2 \, m}
  = - \delta_{m0} T_0 \sqrt{\frac{5\pi}{3}}
       \, \left( \beta^{\rm E}_{k,2} - \beta^{\rm E}_{k,0} \right),
\end{equation}
 where
\begin{align}
 \beta^{\rm E}_{k0} &\simeq \int_0^{\eta_0} d\eta' \, g(\eta') \, G^S_k(\eta') \,
  \left( \frac{2}{3} j_0(k(\eta'-\eta_0)) + \frac{2}{3} j_2(k(\eta'-\eta_0)) \right)
\nonumber\\
 & \simeq \frac{2}{3} \int_0^{\eta_0} d\eta' \, g(\eta') \, G^S_k(\eta') \, j_0(k(\eta'-\eta_0))
\\
 \beta^{\rm E}_{k2} &\simeq - \int_0^{\eta_0} d\eta' \, g(\eta') \, G^S_k(\eta') \,
  \left( \frac{2}{15} j_0(k(\eta'-\eta_0)) + \frac{10}{21} j_2(k(\eta'-\eta_0))
         + \frac{12}{35} j_4(k(\eta'-\eta_0)) \right)
\nonumber\\
 & \simeq - \frac{2}{15} \int_0^{\eta_0} d\eta' \, g(\eta') \, G^S_k(\eta') \, j_0(k(\eta'-\eta_0)).
\end{align}
Here, we have made an approximation following the physical meaning of the integral.
Then,
\begin{equation}
 a^{\rm E}_{k,2 \, m}
  \simeq \delta_{m0} T_0 \frac{4}{5} \sqrt{\frac{5\pi}{3}}
         \int_0^{\eta_0} d\eta' \, g(\eta') \, G^S_k(\eta') \, j_0(k(\eta'-\eta_0))
\end{equation}
 and
\begin{equation}
 C_2^{\rm EE}
  \simeq T_0^2 A_S \frac{16\pi}{75} \int \frac{d^3k}{(2\pi)^3}
         \left( \int_0^{\eta_0} d\eta' \, g(\eta') \, {\tilde G}_k(\eta') \, j_0(k(\eta'-\eta_0)) \right)^2,
\label{power-approx-E}
\end{equation}
 where ${\tilde G}_k(\eta') \equiv G^S_k(\eta')/\sqrt{A_S} = G^T_k(\eta')/\sqrt{A_T}$.

Next we estimate $C_2^{\rm BB}$ in the same way.
The corresponding amplitude is
\begin{equation}
 a^{\rm B}_{k,2 \, m}
  = i ( \delta_{m \, 2} - \delta_{m \, -2}) 
       \frac{T_0}{4} \sqrt{\frac{2\pi}{5}}
       \, \left( 4\beta^{\rm B}_{k,1} + \beta^{\rm B}_{k,3} \right),
\end{equation}
 where
\begin{align}
 \beta^{\rm B}_{k1} &\simeq i \int_0^{\eta_0} d\eta' \, g(\eta') \, G^T_k(\eta') \, j_1(k(\eta'-\eta_0)),
\\
 \beta^{\rm B}_{k3} &\simeq -i \int_0^{\eta_0} d\eta' \, g(\eta') \, G^T_k(\eta') \, j_3(k(\eta'-\eta_0)).
\end{align}
Then,
\begin{align}
 a^{\rm B}_{k,2 \, m}
  & \simeq - ( \delta_{m \, 2} - \delta_{m \, -2}) \frac{T_0}{4} \sqrt{\frac{2\pi}{5}}
      \int_0^{\eta_0} d\eta' \, g(\eta') \, G^T_k(\eta')
      \left( 4 j_1(k(\eta'-\eta_0)) - j_3(k(\eta'-\eta_0)) \right)
\nonumber\\
 & \simeq - ( \delta_{m \, 2} - \delta_{m \, -2}) T_0 \sqrt{\frac{2\pi}{5}}
      \int_0^{\eta_0} d\eta' \, g(\eta') \, G^T_k(\eta') \, j_1(k(\eta'-\eta_0)),
\end{align}
 where we have made the same approximation as before.
Therefore,
\begin{equation}
 C_2^{\rm BB}
  \simeq T_0^2 A_T \frac{16\pi}{50} \int \frac{d^3k}{(2\pi)^3}
         \left( \int_0^{\eta_0} d\eta' \, g(\eta') \, {\tilde G}_k(\eta') \, j_1(k(\eta'-\eta_0)) \right)^2,
\label{power-approx-B}
\end{equation}
 which has very similar form to eq.~(\ref{power-approx-E}).

Now we estimate the integrals of eqs.~(\ref{power-approx-E}) and (\ref{power-approx-B}).
According to the physical meaning of the integral over conformal time
 we replace $k$ of ${\tilde G}_k$ by corresponding typical values of $k^{\rm E}_*$ and $k^{\rm B}_*$
 in eqs.~(\ref{power-approx-E}) and (\ref{power-approx-B}), respectively.
These values are not the same and $k^{\rm B}_*$ is larger then $k^{\rm E}_*$,
 because the order of the corresponding spherical Bessel functions are different, and we estimate
\begin{equation}
 k^{\rm B}_* \simeq \frac{x_1}{x_0} k^{\rm E}_* \simeq \frac{4}{3} k^{\rm E}_*,
\end{equation}
 where $x_0$ and $x_1$ give the first zeros of $j_0(x)$ and $j_1(x)$ at $x \ne 0$, respectively.
Then, eqs.~(\ref{power-approx-E}) becomes
\begin{align}
 C_2^{\rm EE}
  &\simeq T_0^2 A_S \frac{16\pi}{75} \int \frac{d^3k}{(2\pi)^3}
         \left( \frac{1}{300} \sqrt{2 \pi^2 k^{\rm E}_*} \right)^2
\nonumber\\
       & \times
         \int_0^{\eta_0} d\eta' \, g(\eta') \, (\eta'^2-\eta_{\rm ion}^2) \, j_0(k(\eta'-\eta_0))
         \int_0^{\eta_0} d\eta'' \, g(\eta'') \, (\eta''^2-\eta_{\rm ion}^2) \, j_0(k(\eta''-\eta_0)),
\nonumber\\
  & = T_0^2 A_S \frac{16\pi}{75} \left( \frac{1}{300} \sqrt{2 \pi^2 k^{\rm E}_*} \right)^2
      \int_0^{\eta_0} d\eta' d\eta'' g(\eta') (\eta'^2-\eta_{\rm ion}^2) g(\eta'') (\eta''^2-\eta_{\rm ion}^2)
\nonumber\\
  & \times
      \int_0^\infty \frac{4\pi}{(2\pi)^3} dk k^2 j_0(k(\eta'-\eta_0)) j_0(k(\eta''-\eta_0)).
\end{align}
By using the formula of spherical Bessel function
\begin{equation}
 \int_0^\infty  dk k^2 j_\ell(ka) j_\ell(kb) = \frac{\pi}{2ab} \delta(a-b),
\end{equation}
 for $a,b \ne 0$, we obtain
\begin{equation}
 C_2^{\rm EE} \simeq
  T_0^2 A_S \frac{16\pi}{75} \frac{4\pi}{(2\pi)^3} \left( \frac{\sqrt{2\pi^2}}{300} \right)^2 k^{\rm E}_* \,
   \frac{\pi}{2} \int_0^{\eta_0} d\eta'
   \left (g(\eta') \frac{\eta'^2-\eta_{\rm ion}^2}{\eta'-\eta_0} \right)^2.
\end{equation}
In the same way we obtain
\begin{equation}
 C_2^{\rm BB} \simeq
  T_0^2 A_T \frac{16\pi}{50} \frac{4\pi}{(2\pi)^3} \left( \frac{\sqrt{2\pi^2}}{300} \right)^2 k^{\rm B}_* \,
   \frac{\pi}{2} \int_0^{\eta_0} d\eta'
   \left (g(\eta') \frac{\eta'^2-\eta_{\rm ion}^2}{\eta'-\eta_0} \right)^2.
\end{equation}
We see that the reionization process dependence of $C_2^{\rm EE}$ and $C_2^{\rm BB}$ is almost the same and
\begin{equation}
 \frac{C_2^{\rm BB}}{C_2^{\rm EE}}
  \simeq \frac{A_T}{A_S} \cdot \frac{75}{50} \cdot \frac{k^{\rm B}_*}{k^{\rm E}_*}
  \simeq \frac{A_T}{A_S} \cdot \frac{75}{50} \cdot \frac{4}{3} = \frac{1}{2} \, r
\end{equation}
 with eq.~(\ref{tensor-to-scalar}).
The ratio $C_\ell^{\rm BB}/C_\ell^{\rm EE}$ with $\ell=2$
 is expected to be almost insensitive to the reionization process
 and the value is half of tensor-to-scalar ratio.
In the next section
 we numerically investigate the reionization process dependence of the ratio with $\ell \lesssim 10$
 and also compare the value with $\ell=2$ to tensor-to-scalar ratio.

\section{Numerical analyses}
\label{sec:numerical}

\begin{figure}[t]
\centering
\includegraphics[width=65mm]{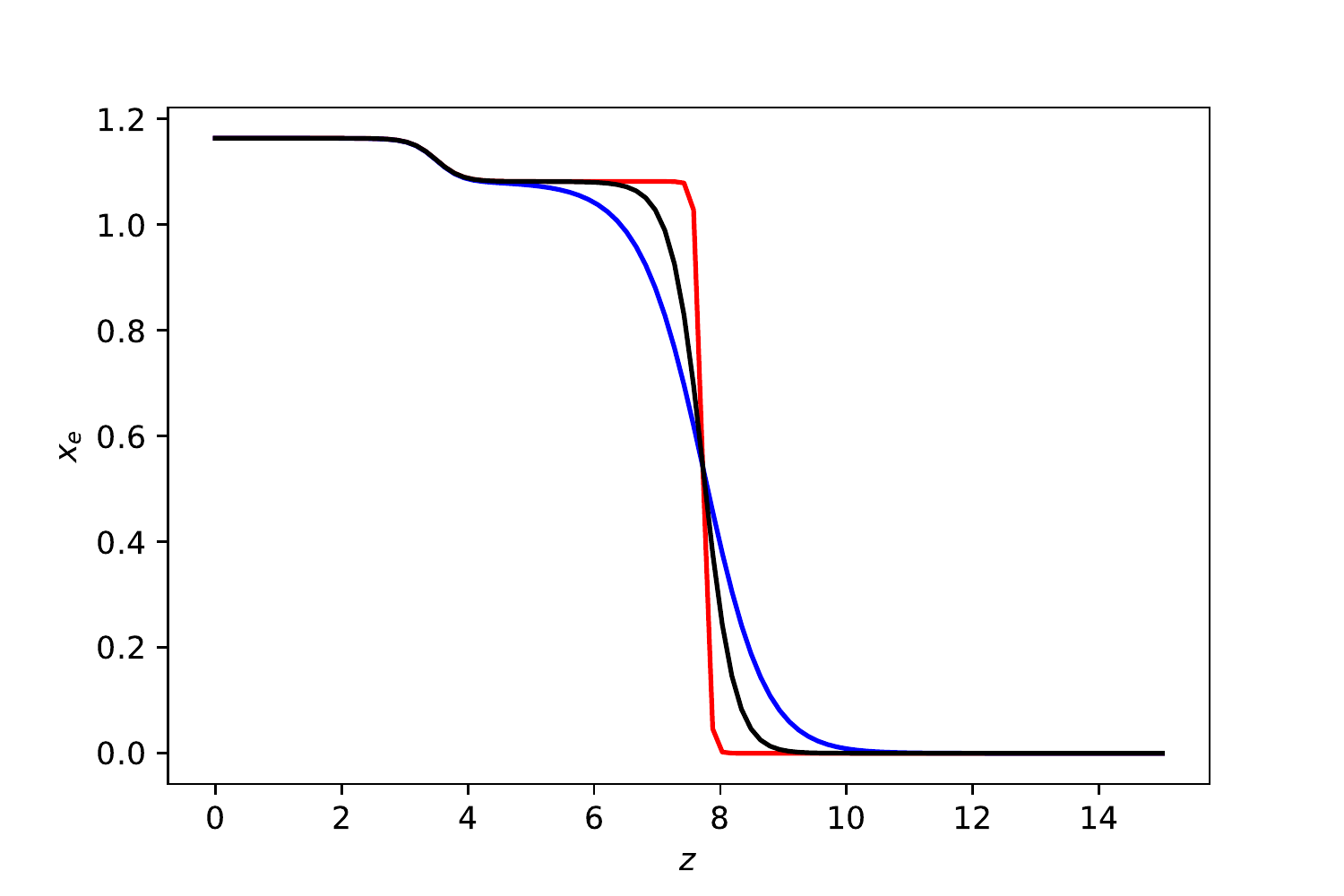}
\quad
\includegraphics[width=65mm]{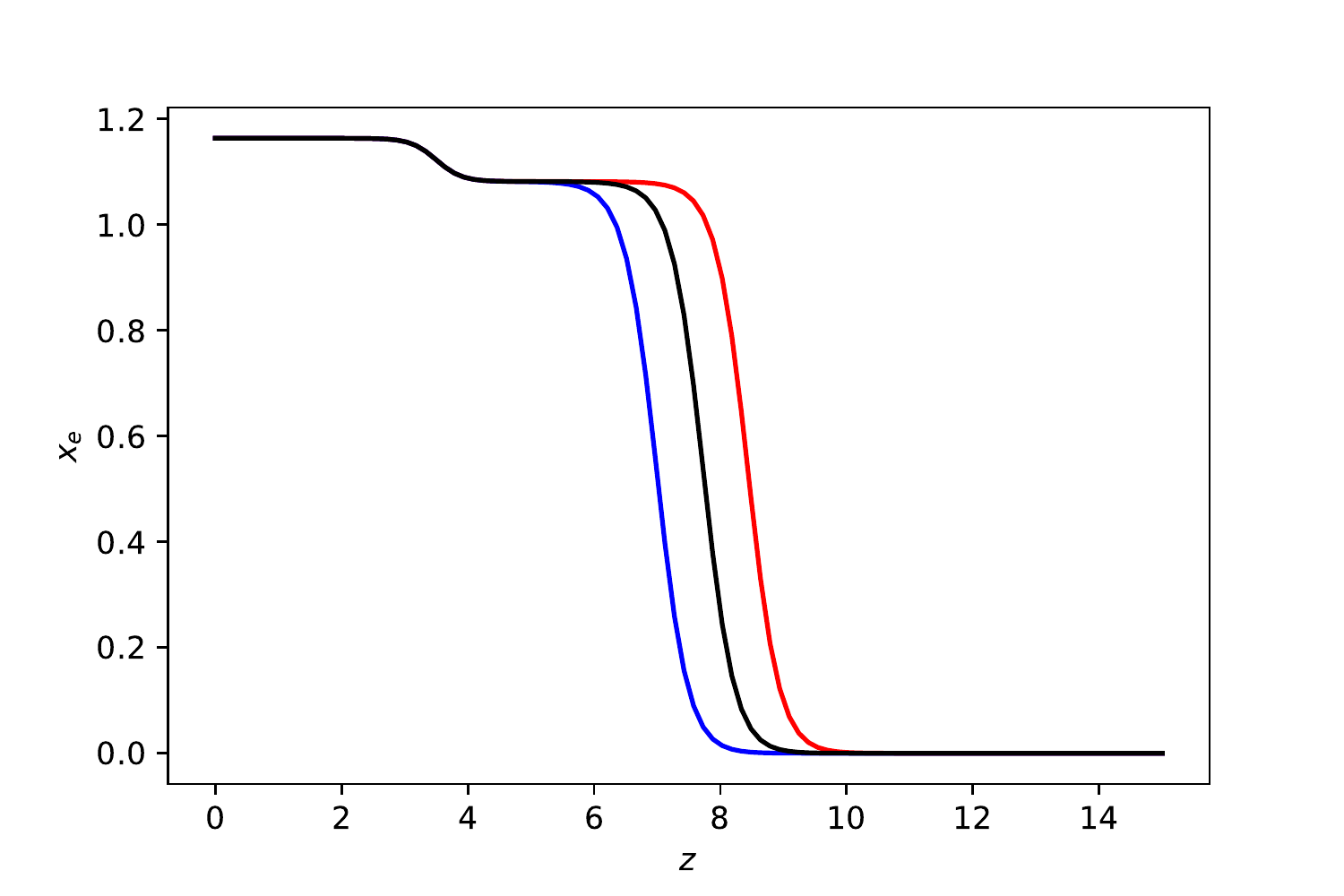}
\caption{
The ionization functions $x_e(z)$.
Left panel: changing $\Delta z_{\rm rei} = 1.0, 0.5$ and $0.1$ with fixed $\tau=0.054$
 corresponding to the lines with blue, black and red, respectively.
Right panel: changing optical depth $\tau = 0.047, 0.054$ and $0.061$ with fixed $\Delta z_{\rm rei} = 0.5$
 corresponding to the lines with blue, black and red, respectively.
}
\label{fig:ionization-functions}
\end{figure}

We use CAMB code \cite{Lewis:1999bs} to investigate reionization process dependence
 of the ratio $C_\ell^{\rm BB}/C_\ell^{\rm EE}$ with $\ell \lesssim 10$.
The $\Lambda$CDM model is assumed with the following values of cosmological parameters:
 $\Omega_b h^2 = 0.022$, $\Omega_m h^2 = 0.12$, $A_S = 2\times10^{-9}$, $n_s = 0.965$
 and $H_0=67.4$ which are obtained by the Planck collaboration \cite{Planck:2018vyg}.
For tensor-to-scalar ratio
 three typical values of $r=0.03, 0.01$ and $0.003$ are considered.

\begin{figure}[t]
\centering
\includegraphics[width=50mm]{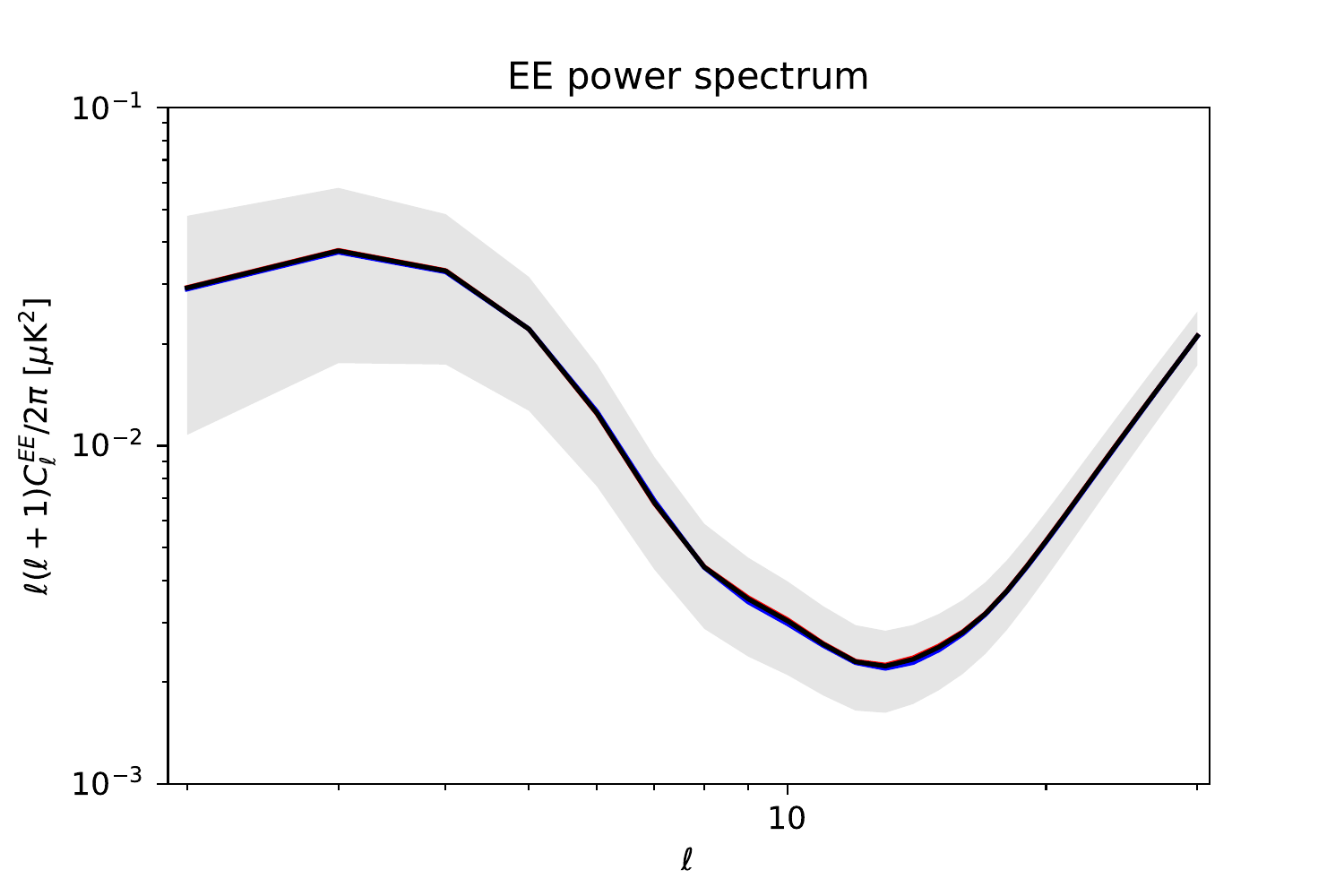}
\quad
\includegraphics[width=50mm]{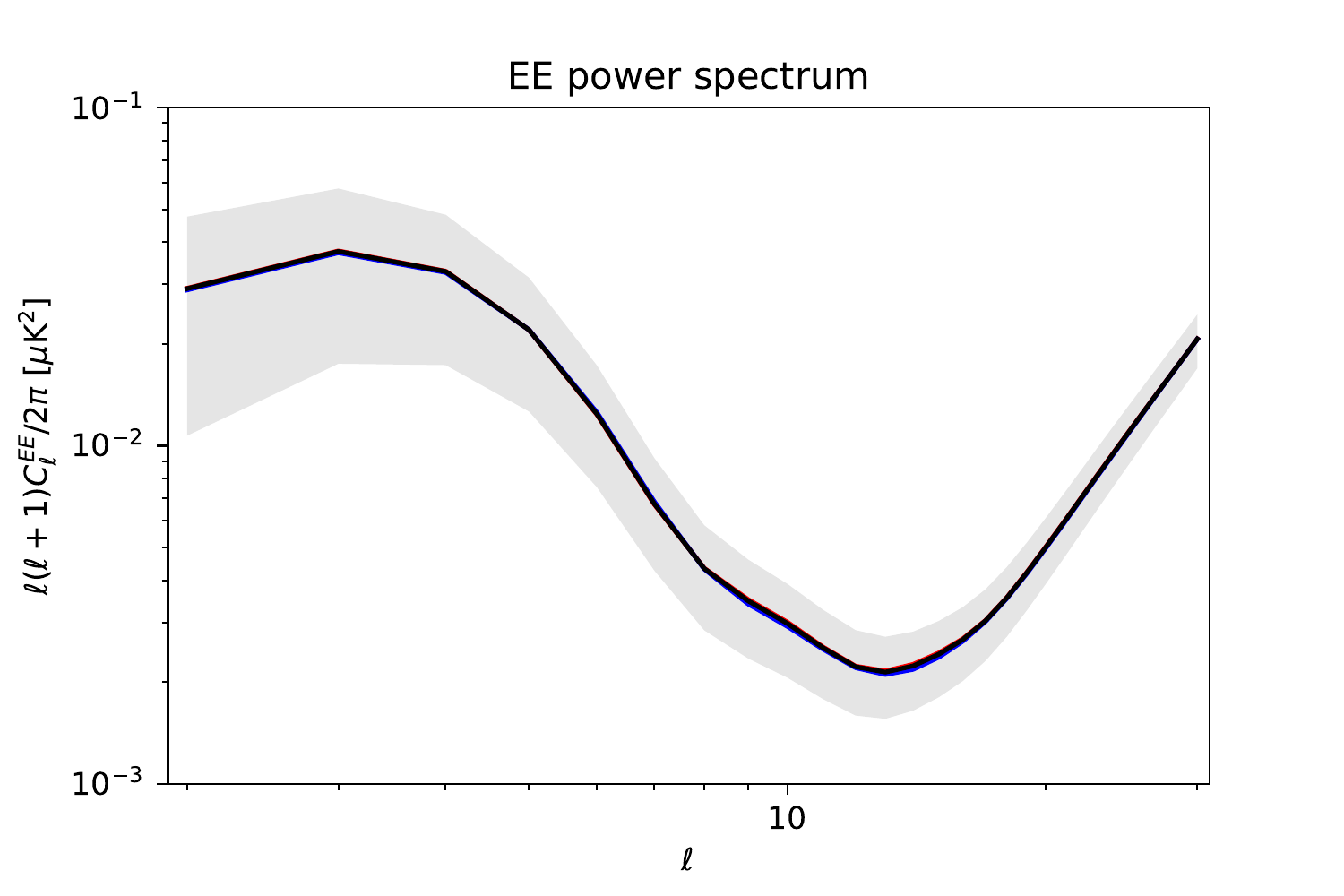}
\quad
\includegraphics[width=50mm]{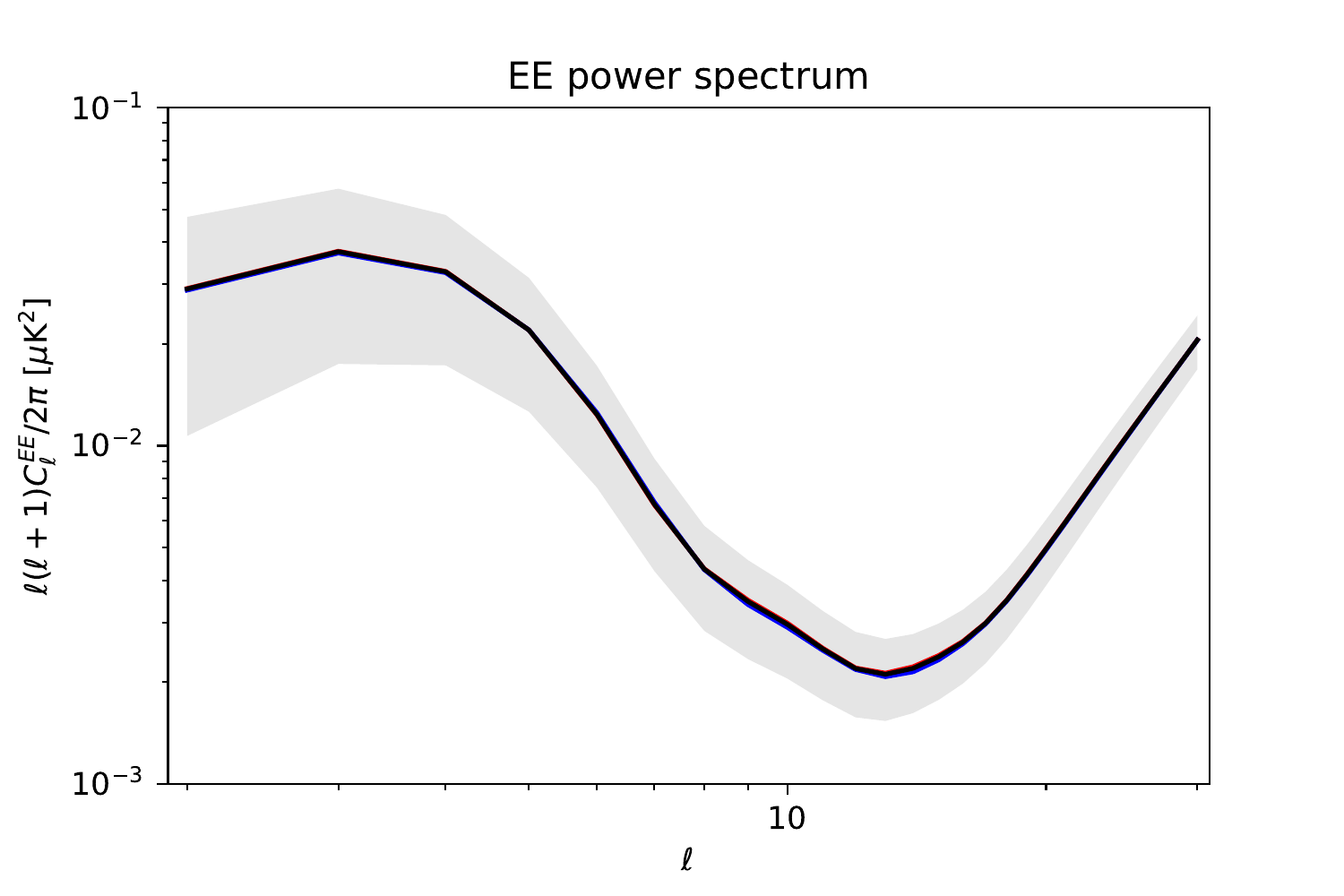}
\caption{
EE power spectra changing $\Delta z_{\rm rei}$.
The same color conventions as in the left panel of Fig.~\ref{fig:ionization-functions}.
Three lines almost overlap.
The value of $r=0.03, 0.01$ and $0.003$ from left to right.
The shaded areas indicate the cosmic variance corresponding to the black line.
}
\label{fig:EE-power-changing-slope}
\end{figure}
\begin{figure}[t]
\centering
\includegraphics[width=50mm]{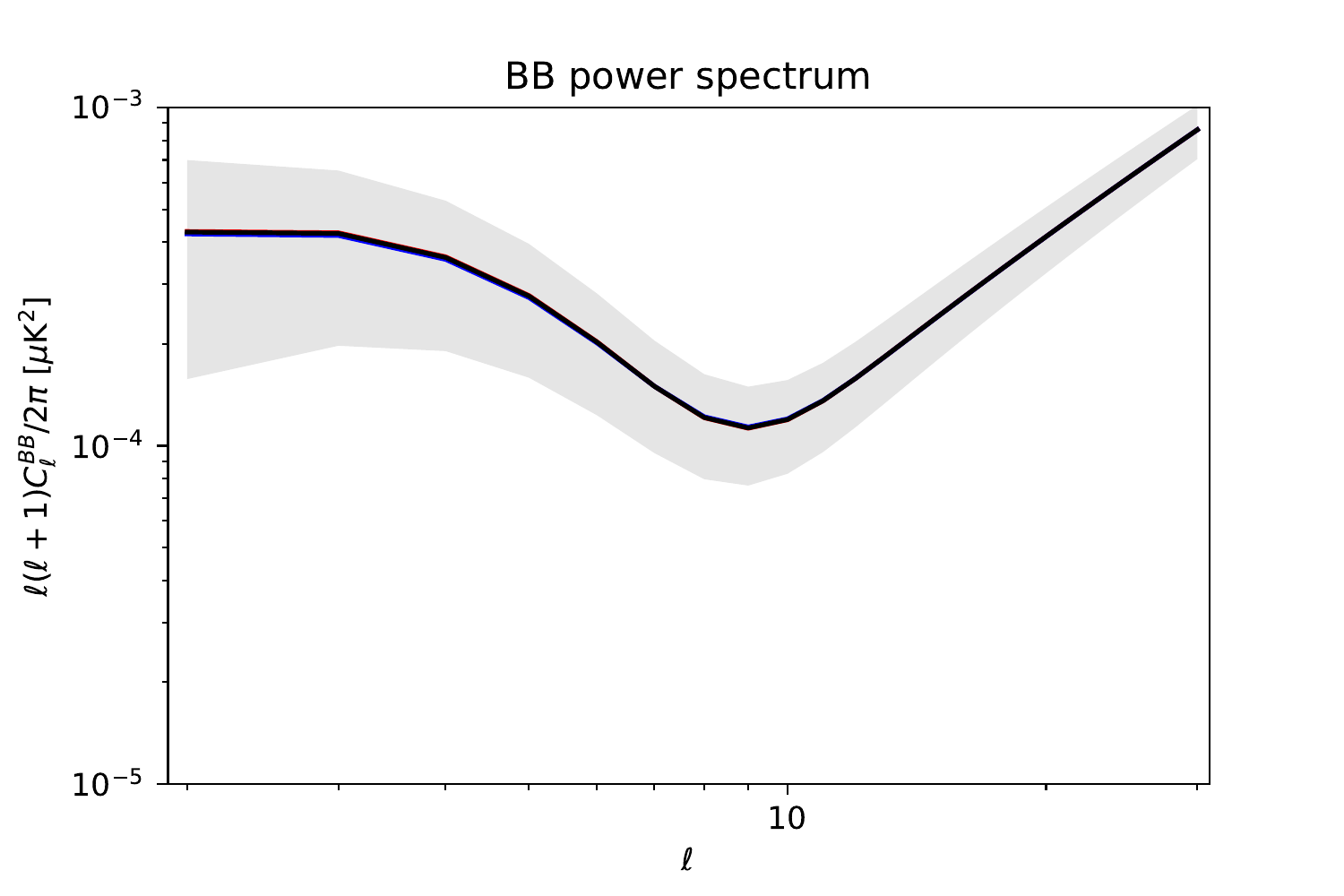}
\quad
\includegraphics[width=50mm]{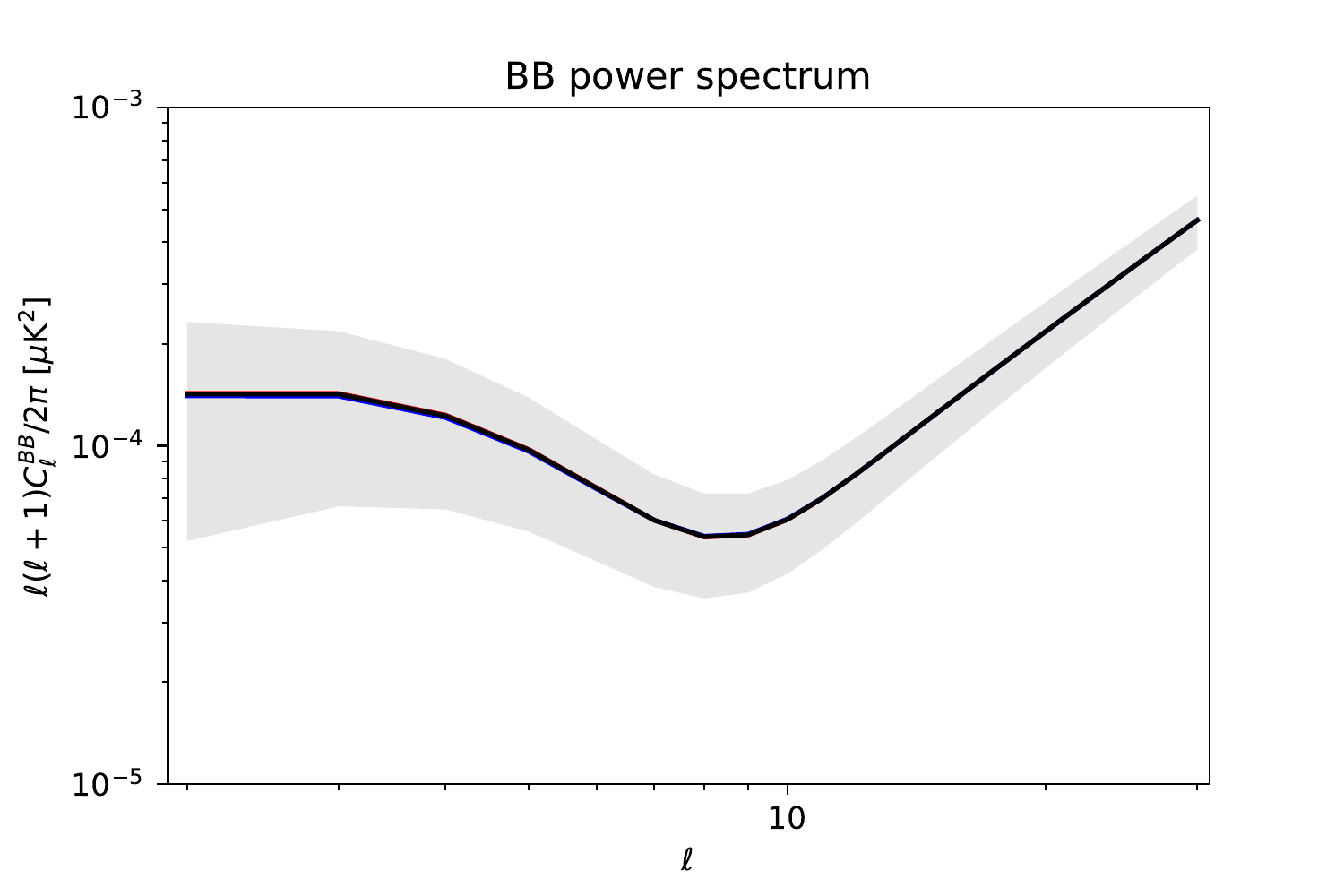}
\quad
\includegraphics[width=50mm]{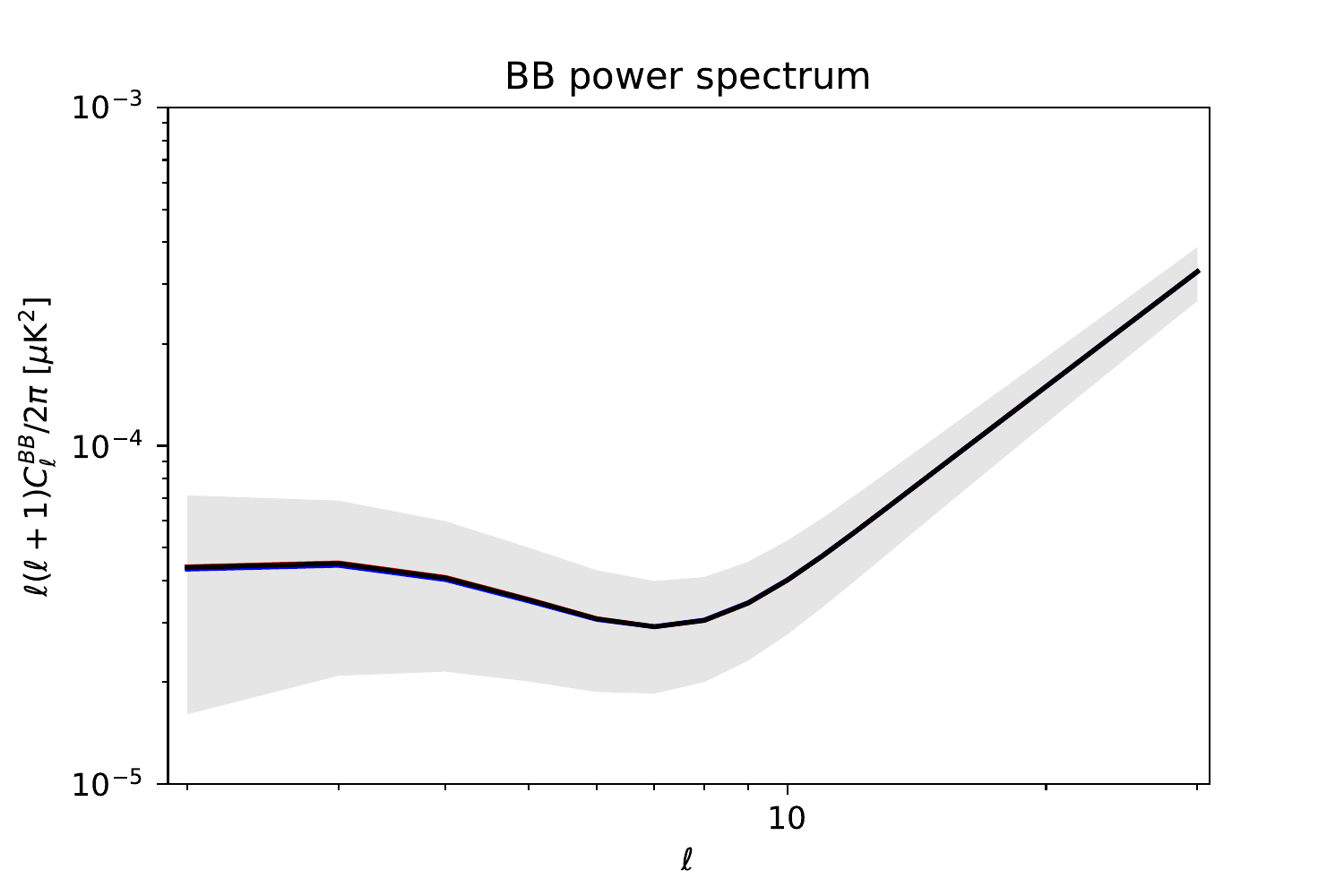}
\caption{
BB power spectra changing $\Delta z_{\rm rei}$.
The same color conventions as in the left panel of Fig.~\ref{fig:ionization-functions}.
Three lines almost overlap.
The value of $r=0.03, 0.01$ and $0.003$ from left to right.
The shaded areas indicate the cosmic variance corresponding to the black line.
}
\label{fig:BB-power-changing-slope}
\end{figure}

Since the standard reionization theory has not yet established,
 as it has been discussed in the first section,
 we consider several variations of simple ionization function, $x_e(z)$,
 which consists of two $\tanh$ functions including the ionization of second electrons of helium.
Such a simple model is contained in the CAMB code as default,
 which includes only two parameters: optical depth $\tau$ and $\Delta z_{\rm rei}$
 which describes the duration of hydrogen ionization in redshift.
Fig.~\ref{fig:ionization-functions}
 shows six ionization functions which we use in this analysis.
The value of the optical depth $\tau$ is changed
 within the constraint $\tau = 0.054 \pm 0.007$ ($0.047 \leq \tau \leq 0.061$)
 which is obtained by the Planck collaboration \cite{Planck:2018vyg}
 with fixed $\Delta z_{\rm rei} = 0.5$.
The duration of hydrogen ionization $\Delta z_{\rm rei}$
 is changed from slow extreme to fast extreme
 within the constraint that reionization process should be finished until $z \simeq 5.2 \sim 5.6$
 \cite{Choudhury:2020vzu,Qin:2021gkn} with fixed $\tau = 0.054$.
The helium second ionization is considered to be the same for all cases for simplicity
 with the amount of helium $n_{{\rm He}0}/n_{{\rm H}0}=0.08$.
Note that the reionization starts later for smaller $\tau$ or smaller $\Delta z_{\rm rei}$.

\begin{figure}[t]
\centering
\includegraphics[width=50mm]{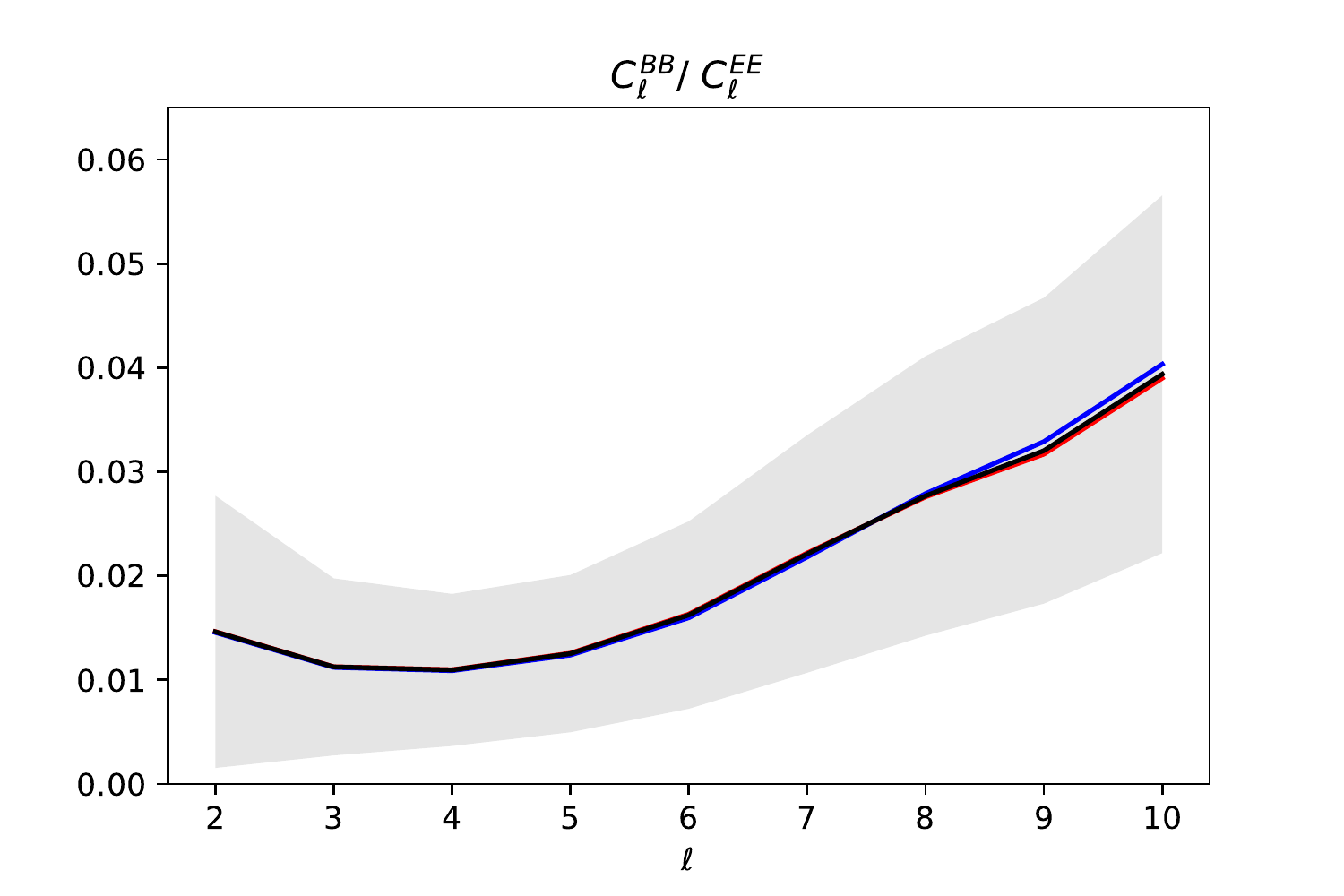}
\quad
\includegraphics[width=50mm]{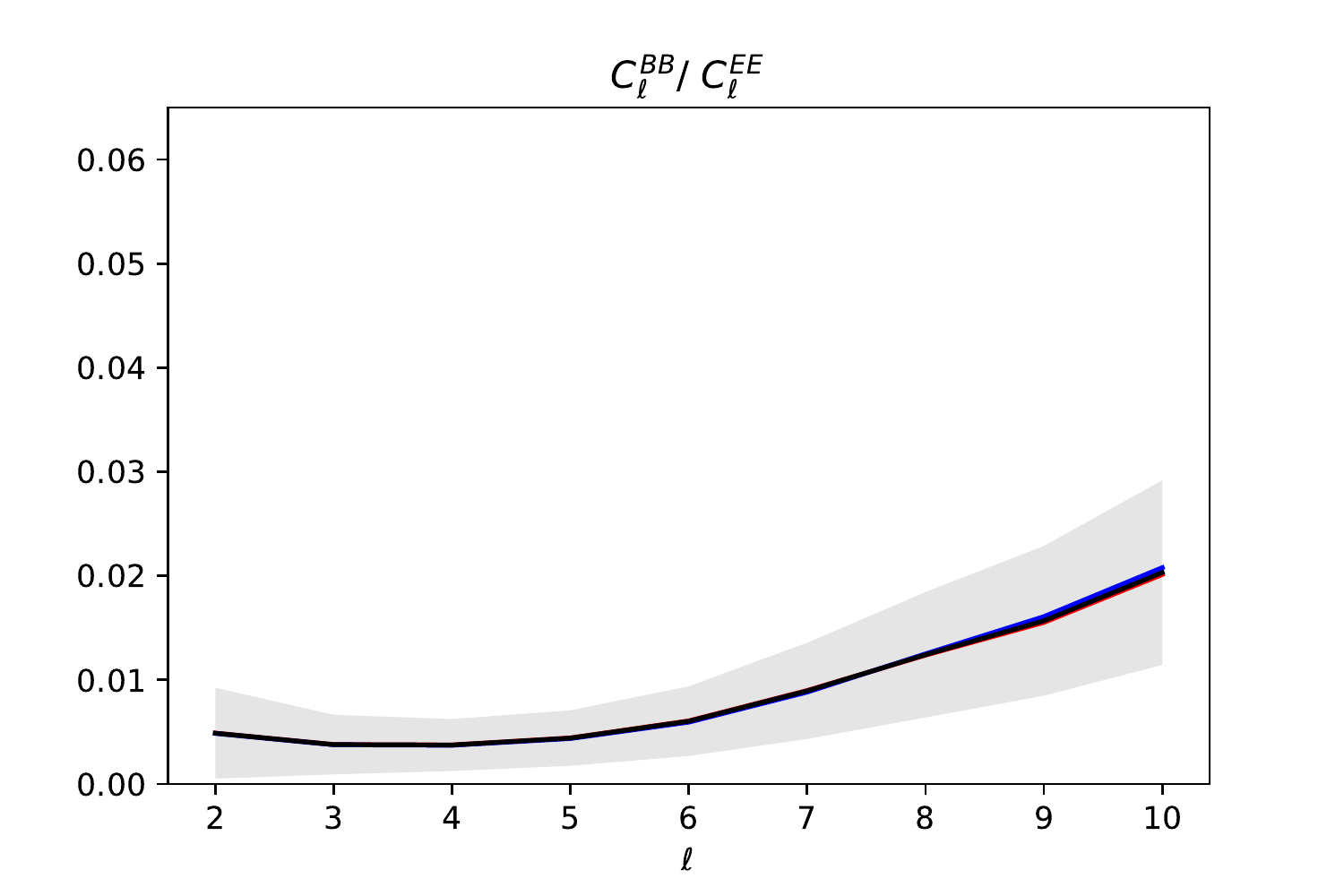}
\quad
\includegraphics[width=50mm]{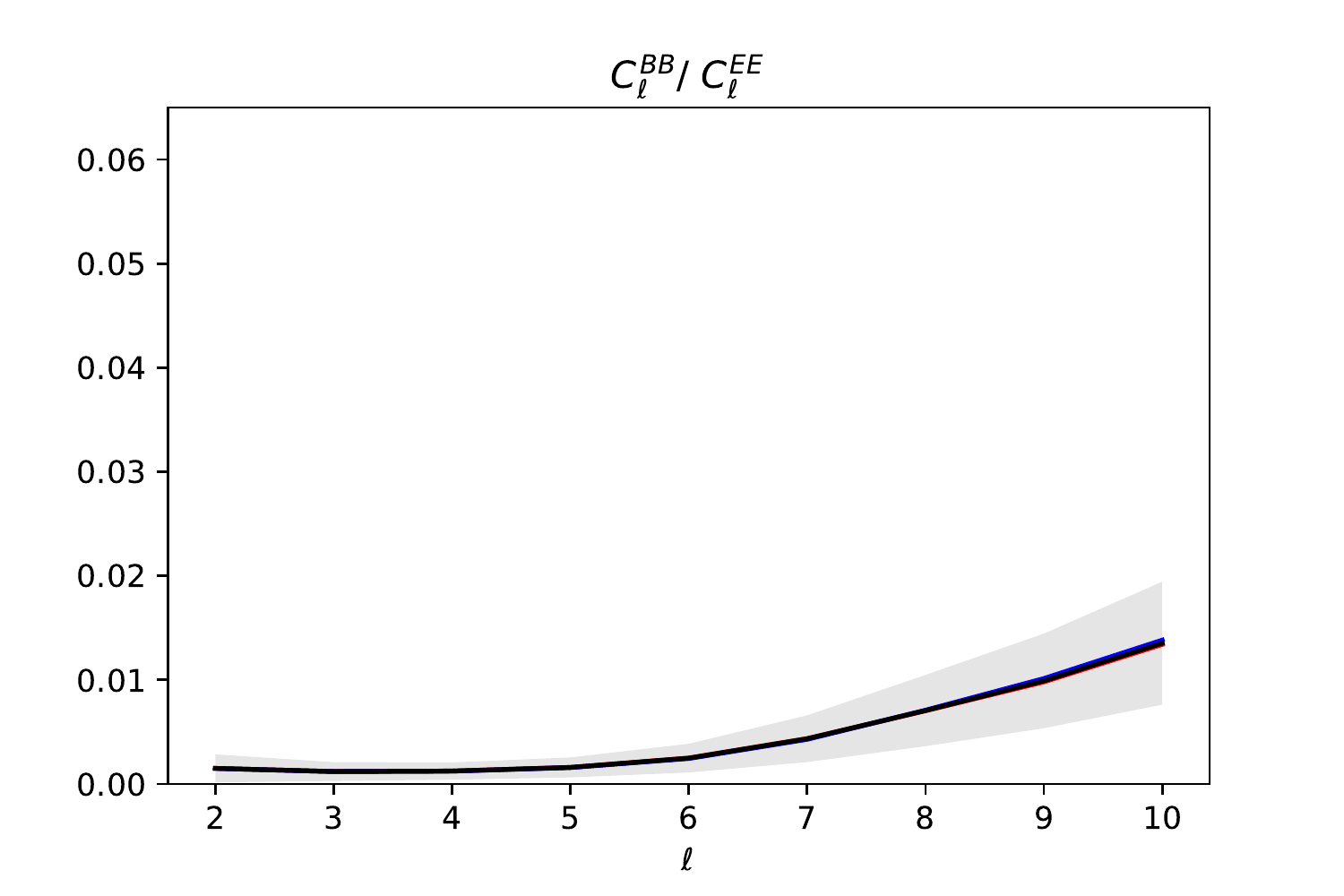}
\caption{
The ratios of $C^{\rm BB}_\ell / C^{\rm EE}_\ell$ changing $\Delta z_{\rm rei}$.
The same color conventions as in the left panel of Fig.~\ref{fig:ionization-functions}.
Three lines almost overlap.
The value of $r=0.03, 0.01$ and $0.003$ from left to right.
The shaded areas indicate the cosmic variance corresponding to the black line.
}
\label{fig:ratio-changing-slope}
\end{figure}
\begin{figure}[t]
\centering
\includegraphics[width=50mm]{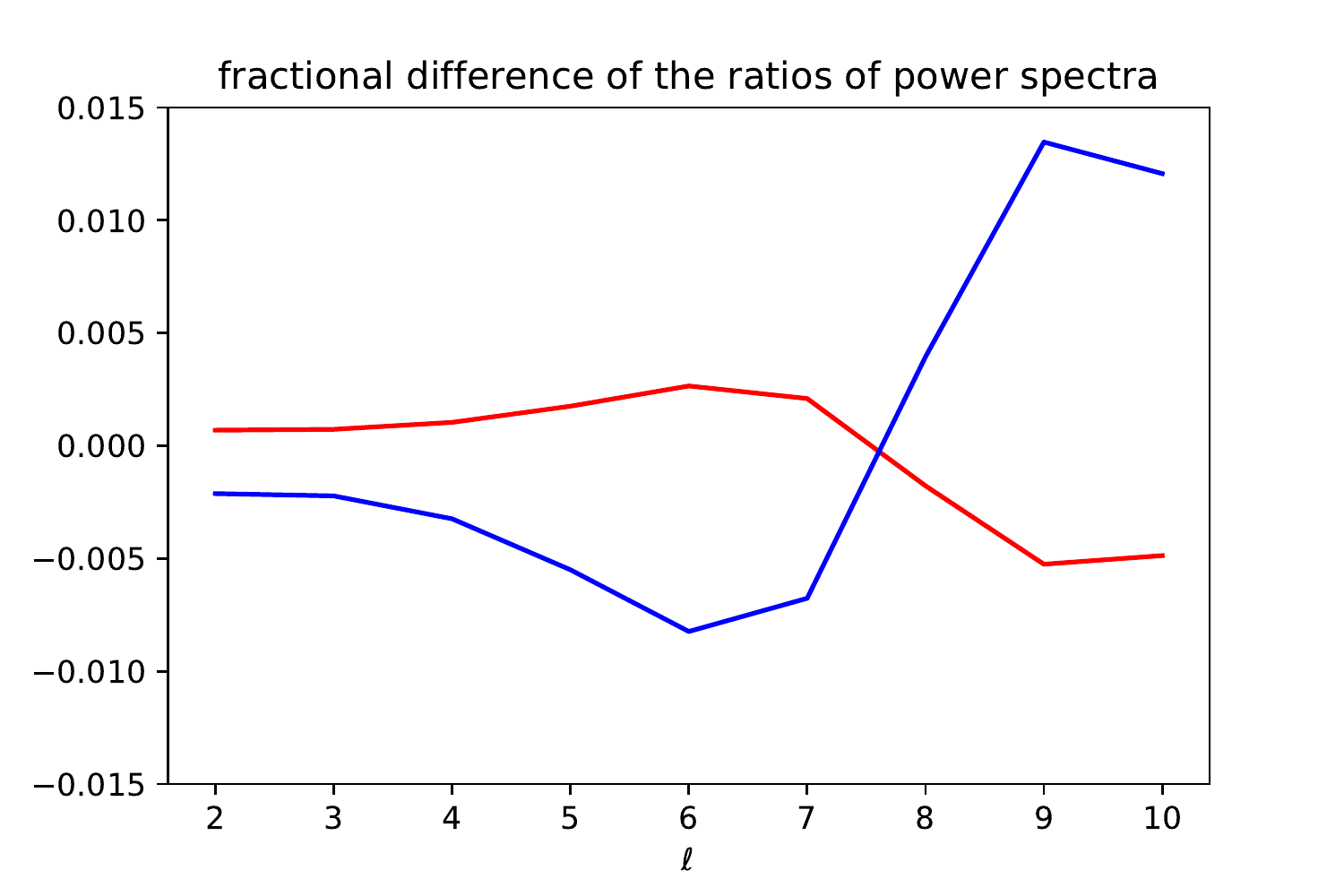}
\quad
\includegraphics[width=50mm]{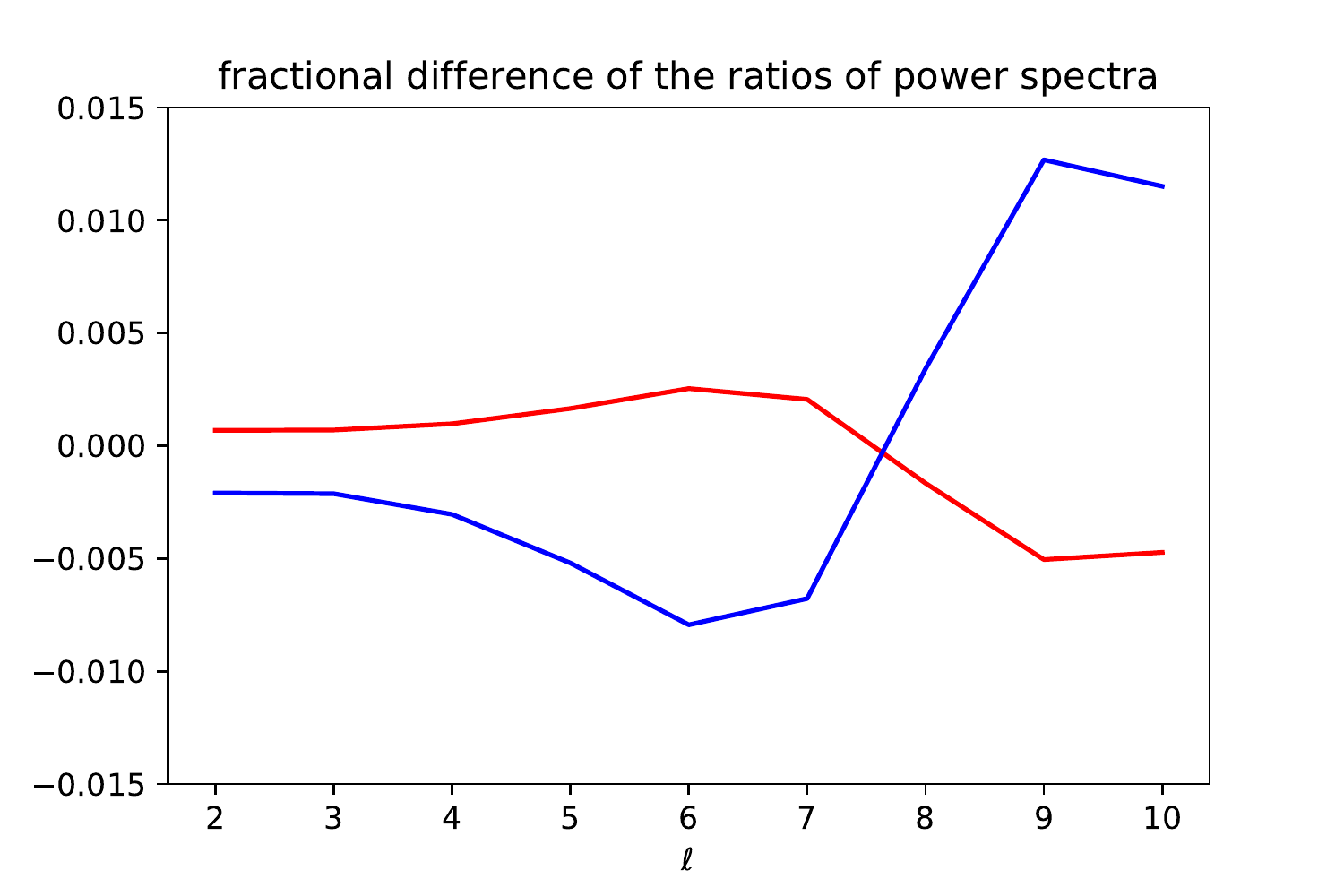}
\quad
\includegraphics[width=50mm]{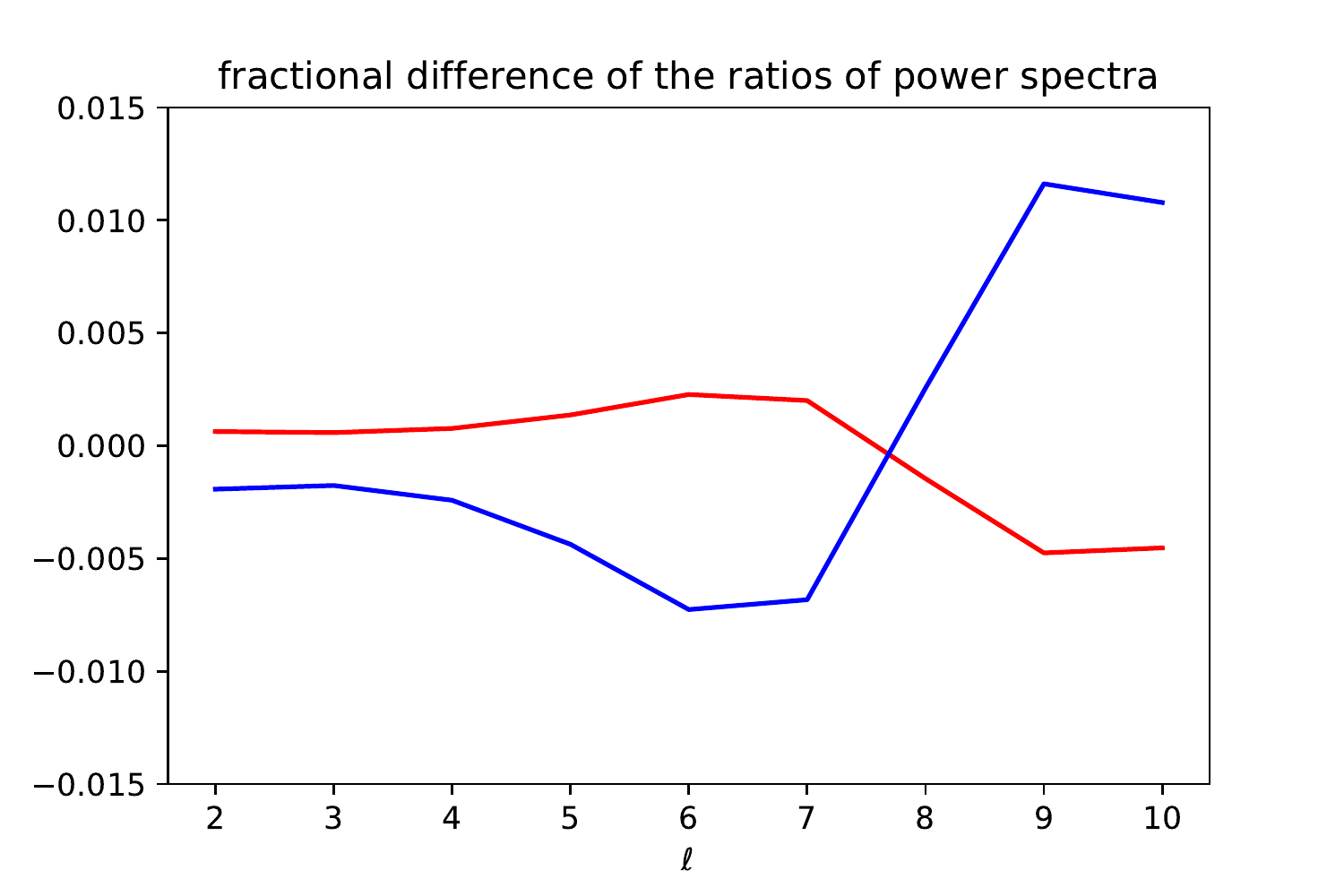}
\caption{
The fractional differences of ratio $C^{\rm BB}_\ell / C^{\rm EE}_\ell$ between
 the case of $\Delta z_{\rm rei} = 1.0$ and $\Delta z_{\rm rei} = 0.5$ (blue),
 and between
 the case of $\Delta z_{\rm rei} = 0.1$ and $\Delta z_{\rm rei} = 0.5$ (red).
The value of $r=0.03, 0.01$ and $0.003$ from left to right.
}
\label{fig:diff-ratio-changing-slope}
\end{figure}

We first consider the case of changing $\Delta z_{\rm rei}$, or the duration of ionization,
 with fixed optical depth $\tau = 0.054$.
Figs.~\ref{fig:EE-power-changing-slope} and \ref{fig:BB-power-changing-slope}
 show the EE and BB power spectra.
We see that
 the differences by changing $\Delta z_{\rm rei}$ are very small,
 totally within cosmic variance, in all values of $r$.
The same applies for the ratio $C^{\rm BB}_\ell / C^{\rm EE}_\ell$
 as in Fig.~\ref{fig:ratio-changing-slope}.
Fig.~\ref{fig:diff-ratio-changing-slope} shows the fractional differences of the ratios
\begin{equation}
 \frac{\left( \frac{C^{\rm BB}_\ell}{C^{\rm EE}_\ell} \right)_{\Delta z_{\rm rei} = 1.0}
       - \left( \frac{C^{\rm BB}_\ell}{C^{\rm EE}_\ell} \right)_{\Delta z_{\rm rei} = 0.5}}
      {\left( \frac{C^{\rm BB}_\ell}{C^{\rm EE}_\ell} \right)_{\Delta z_{\rm rei} = 1.0}
       + \left( \frac{C^{\rm BB}_\ell}{C^{\rm EE}_\ell} \right)_{\Delta z_{\rm rei} = 0.5}}
\quad {\rm and} \quad
 \frac{\left( \frac{C^{\rm BB}_\ell}{C^{\rm EE}_\ell} \right)_{\Delta z_{\rm rei} = 0.1}
       - \left( \frac{C^{\rm BB}_\ell}{C^{\rm EE}_\ell} \right)_{\Delta z_{\rm rei} = 0.5}}
      {\left( \frac{C^{\rm BB}_\ell}{C^{\rm EE}_\ell} \right)_{\Delta z_{\rm rei} = 0.1}
       + \left( \frac{C^{\rm BB}_\ell}{C^{\rm EE}_\ell} \right)_{\Delta z_{\rm rei} = 0.5}}
\end{equation}
 with blue and red lines, respectively.
The $\ell$-dependence of fractional differences is almost independent from the value of $r$.
We find that $\Delta z_{\rm rei}$-dependence
 of the value of ratio $C^{\rm BB}_\ell / C^{\rm EE}_\ell$ is very small
 less than 1.5\%, especially less than 0.5\%  for $\ell < 5$
 in fractional difference with fixed $\tau=0.054$.
In case of $\Delta z_{\rm rei}=1.0$,
 both EE and BB power spectra decrease very little for small $\ell < 5$
 due to the smaller number of free electrons at $z \lesssim 6$,
 and the amount of the change is larger in the BB power than the EE power,
 which explains the shape of blue lines for $\ell < 5$ in Fig.~\ref{fig:diff-ratio-changing-slope}.
The opposite is true in case of $\Delta z_{\rm rei}=0.1$,
 which explains the shape of red lines for $\ell < 5$ in Fig.~\ref{fig:diff-ratio-changing-slope}.
The values of the ratio
 $C^{\rm BB}_\ell / C^{\rm EE}_\ell$ with $\ell=2$ divided by tensor-to-scalar ratio $r$
 are $0.49$, $0.49$ and $0.50$ for $r=0.03, 0.01$ and $0.003$, respectively,
 in accord with the analytical estimate in the previous section.

\begin{figure}[t]
\centering
\includegraphics[width=50mm]{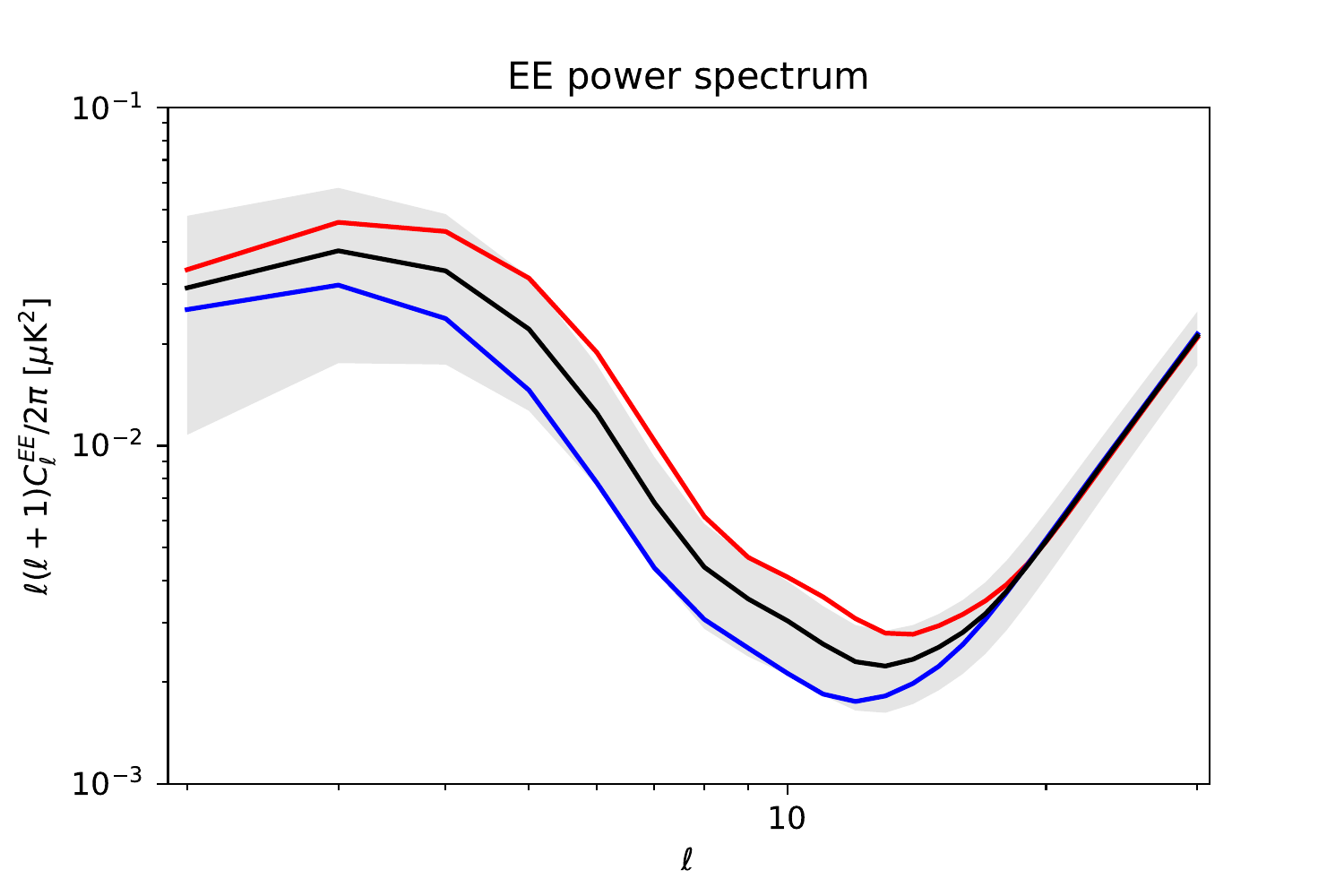}
\quad
\includegraphics[width=50mm]{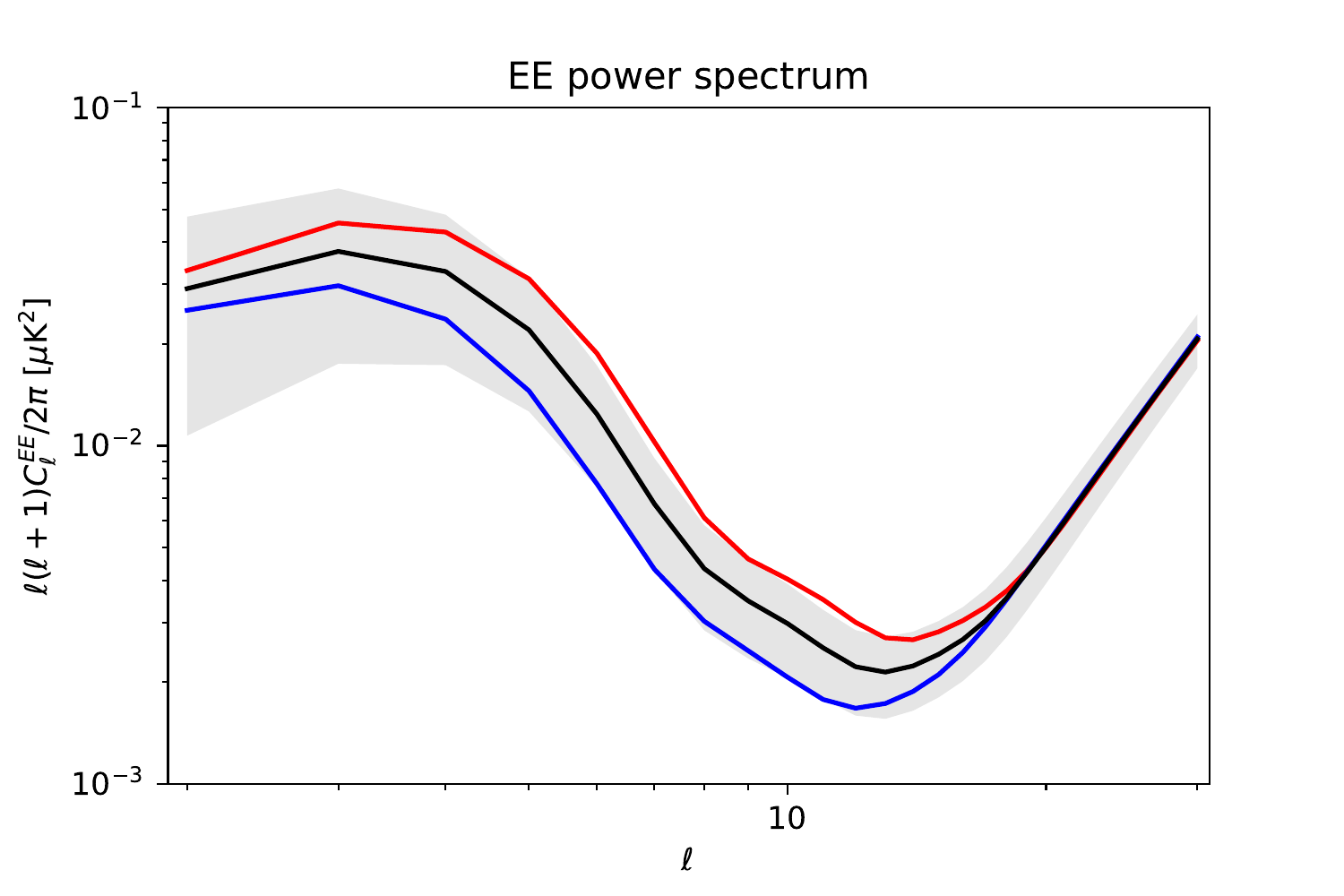}
\quad
\includegraphics[width=50mm]{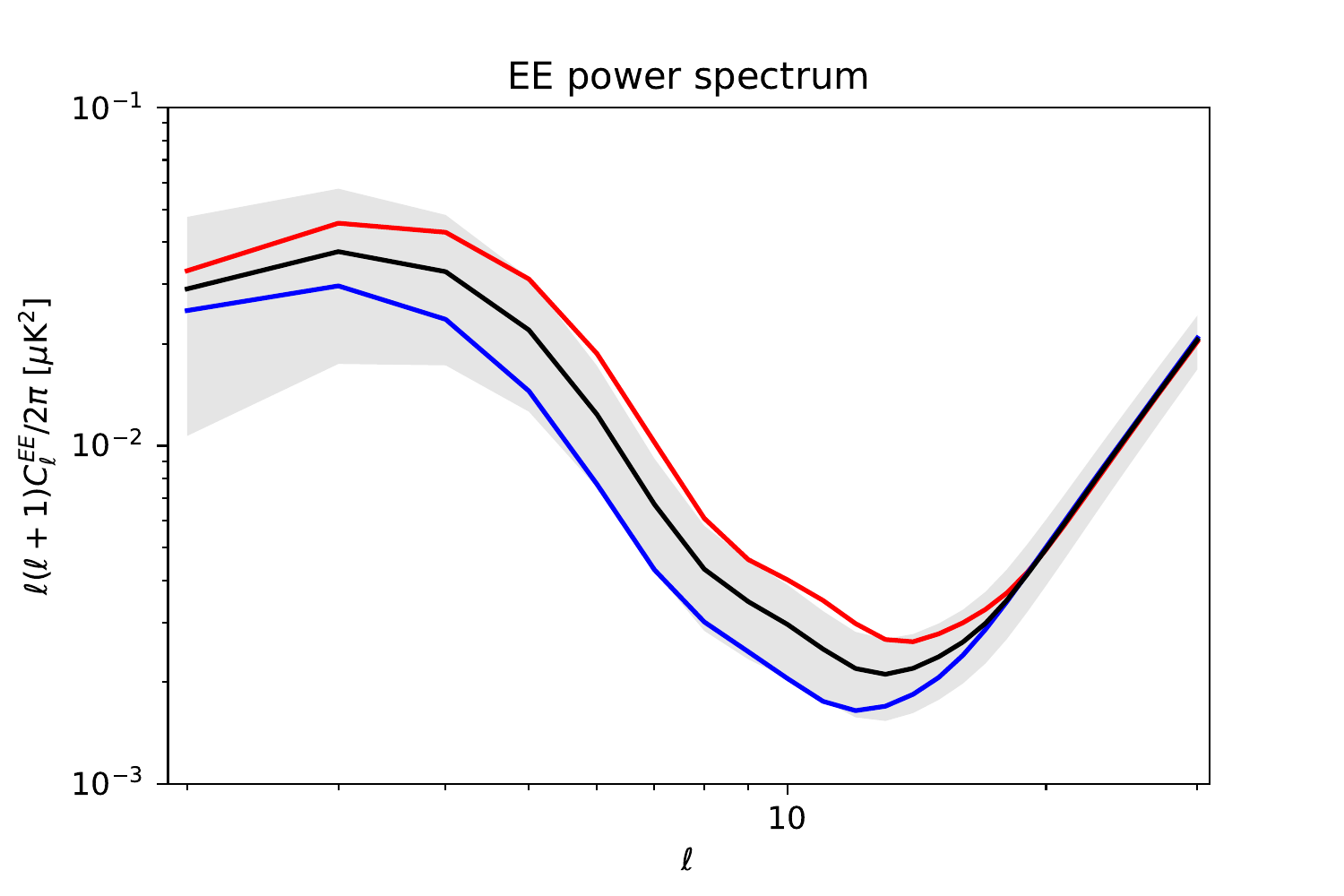}
\caption{
EE power spectra changing optical depth $\tau$.
The same color conventions as in the right panel of Fig.~\ref{fig:ionization-functions}.
The value of $r=0.03, 0.01$ and $0.003$ from left to right.
The shaded areas indicate the cosmic variance corresponding to the black line.
}
\label{fig:EE-power-changing-tau}
\end{figure}
\begin{figure}[t]
\centering
\includegraphics[width=50mm]{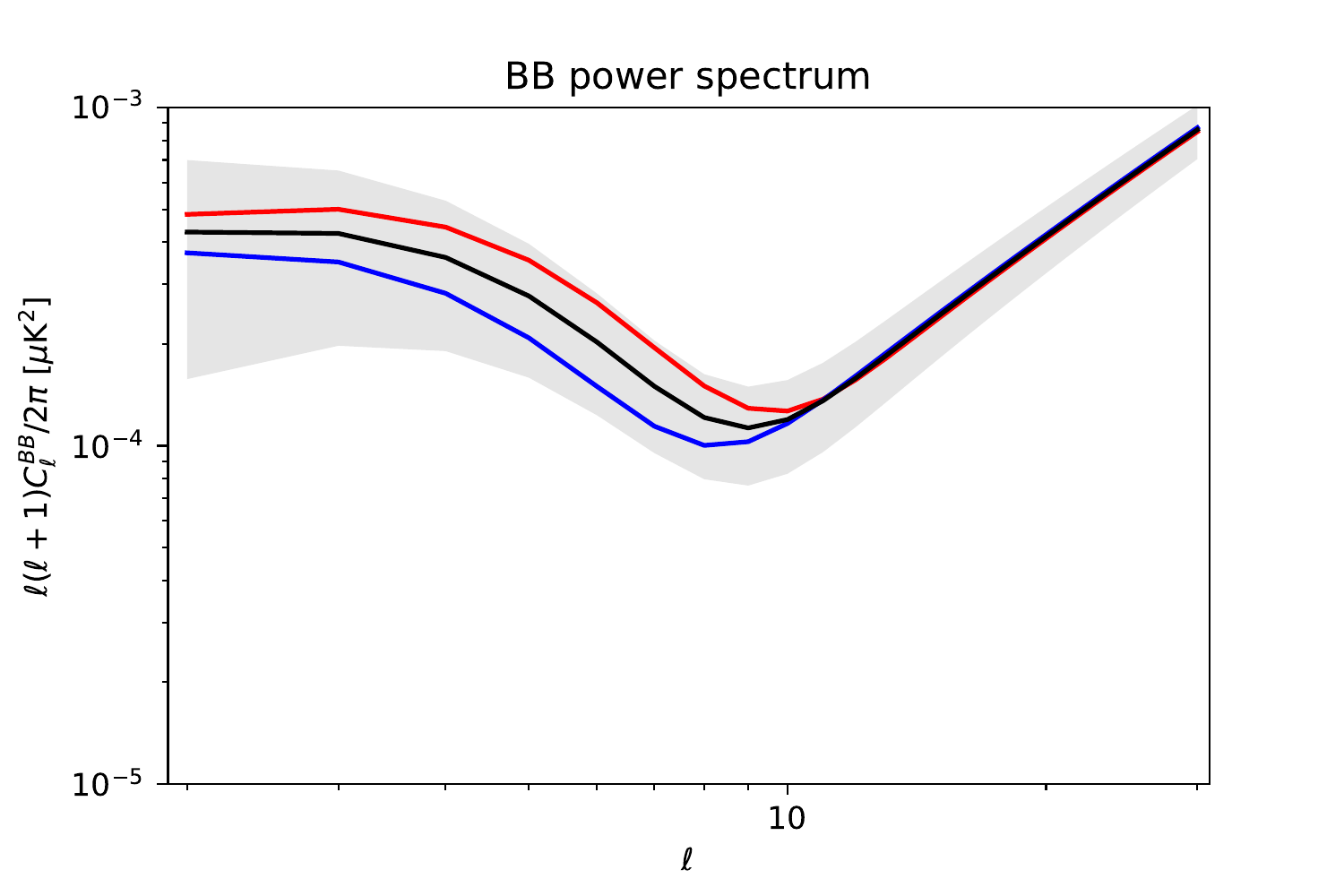}
\quad
\includegraphics[width=50mm]{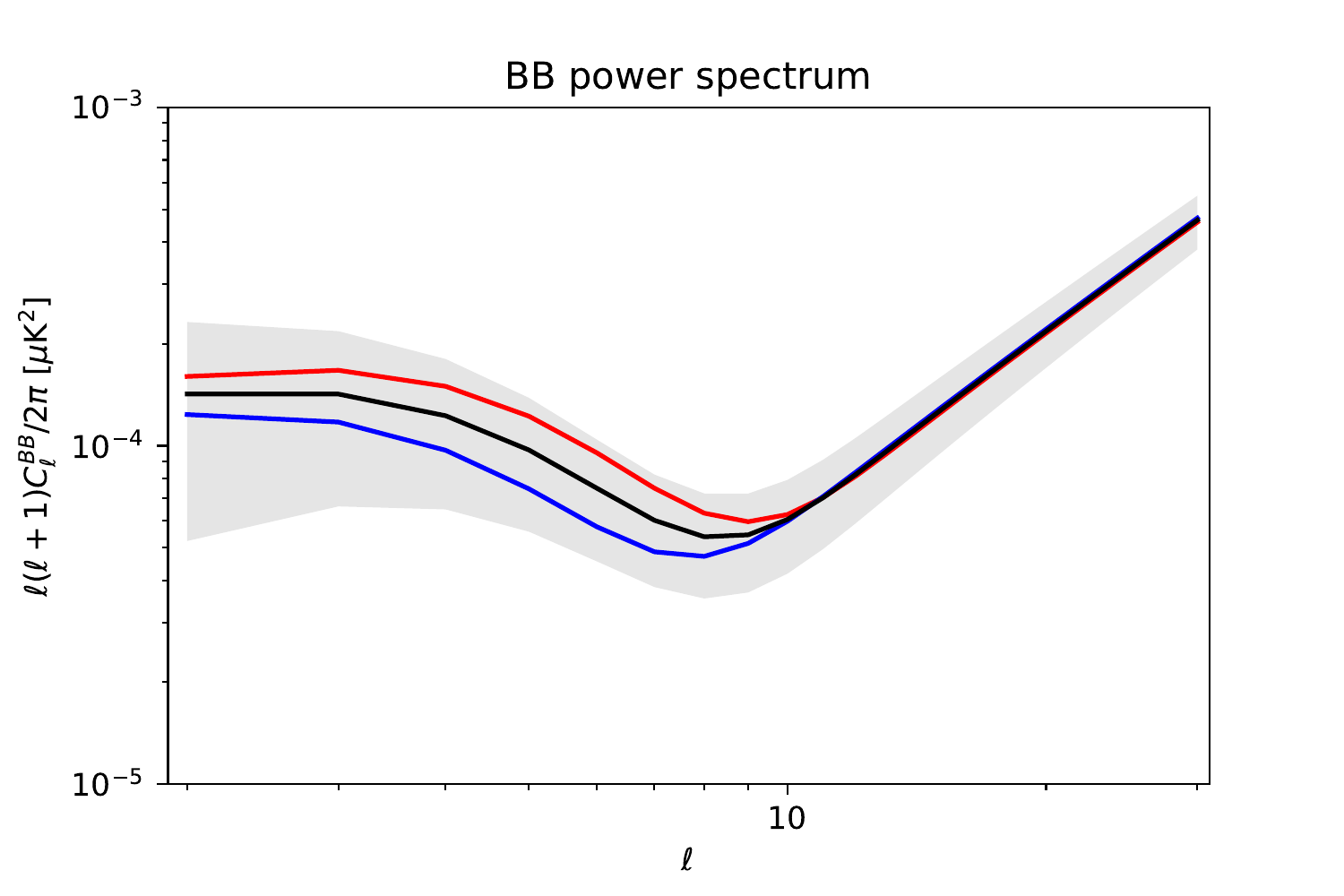}
\quad
\includegraphics[width=50mm]{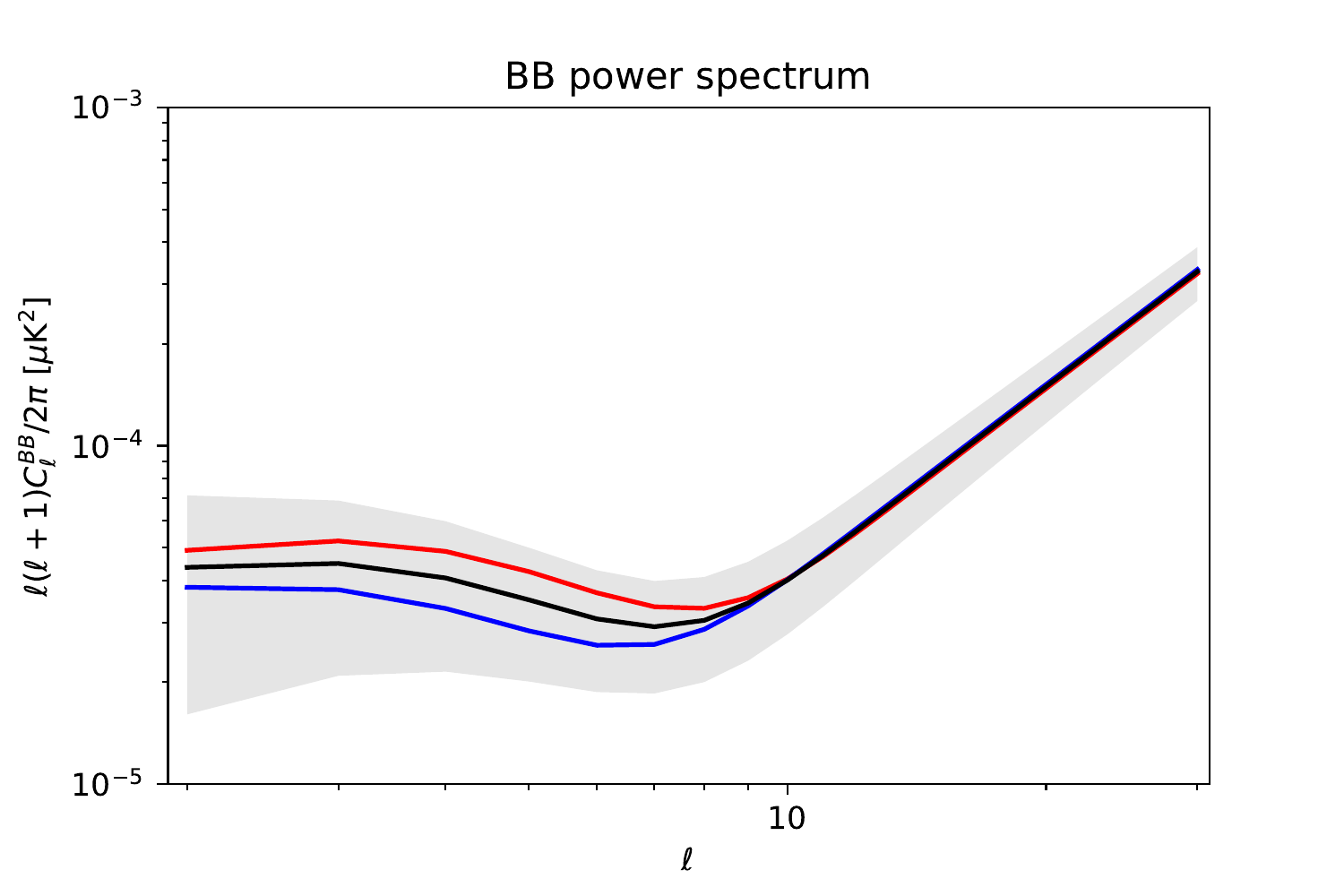}
\caption{
BB power spectra changing optical depth $\tau$.
The same color conventions as in the right panel of Fig.~\ref{fig:ionization-functions}.
The value of $r=0.03, 0.01$ and $0.003$ from left to right.
The shaded areas indicate the cosmic variance corresponding to the black line.
}
\label{fig:BB-power-changing-tau}
\end{figure}

\begin{figure}[t]
\centering
\includegraphics[width=50mm]{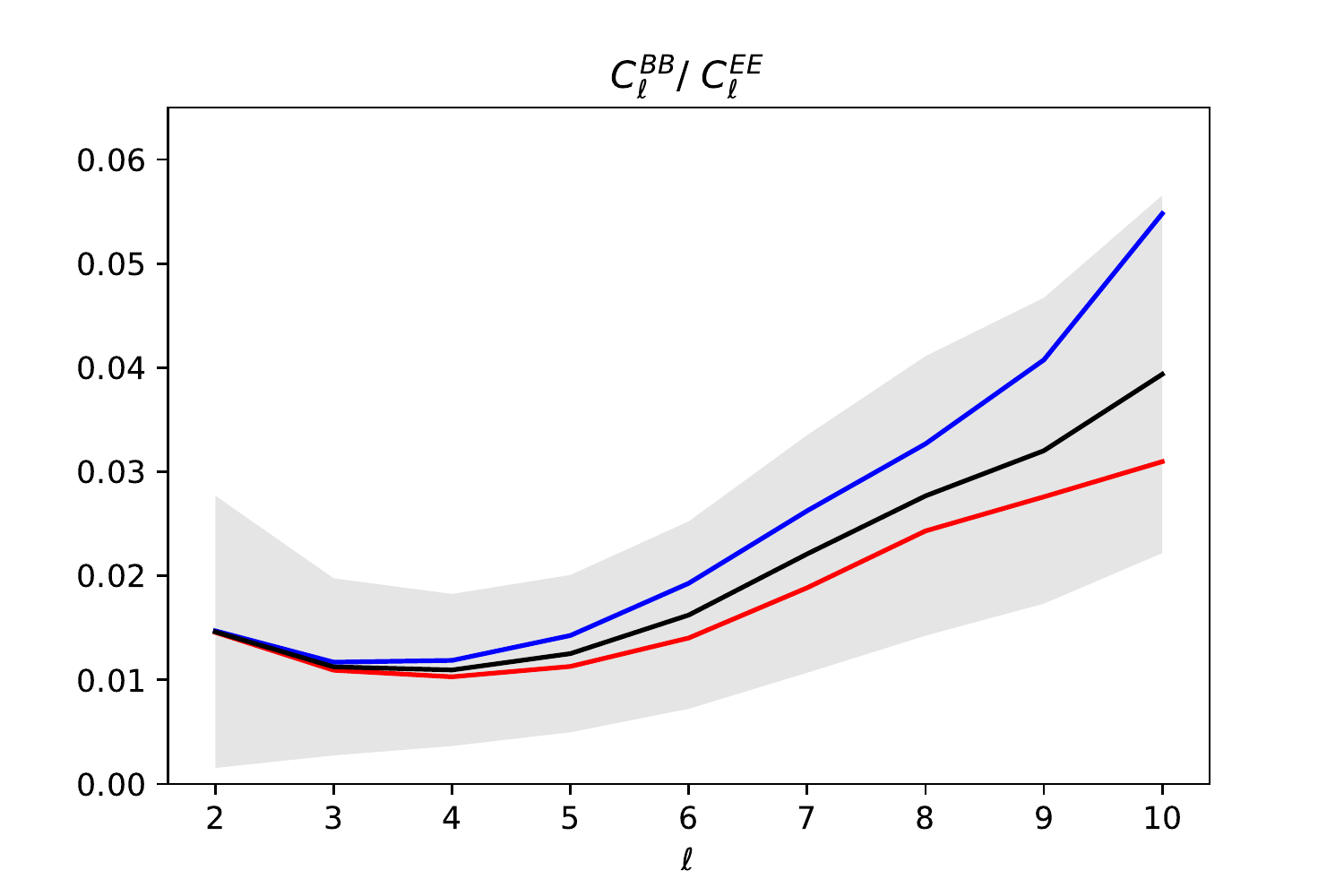}
\quad
\includegraphics[width=50mm]{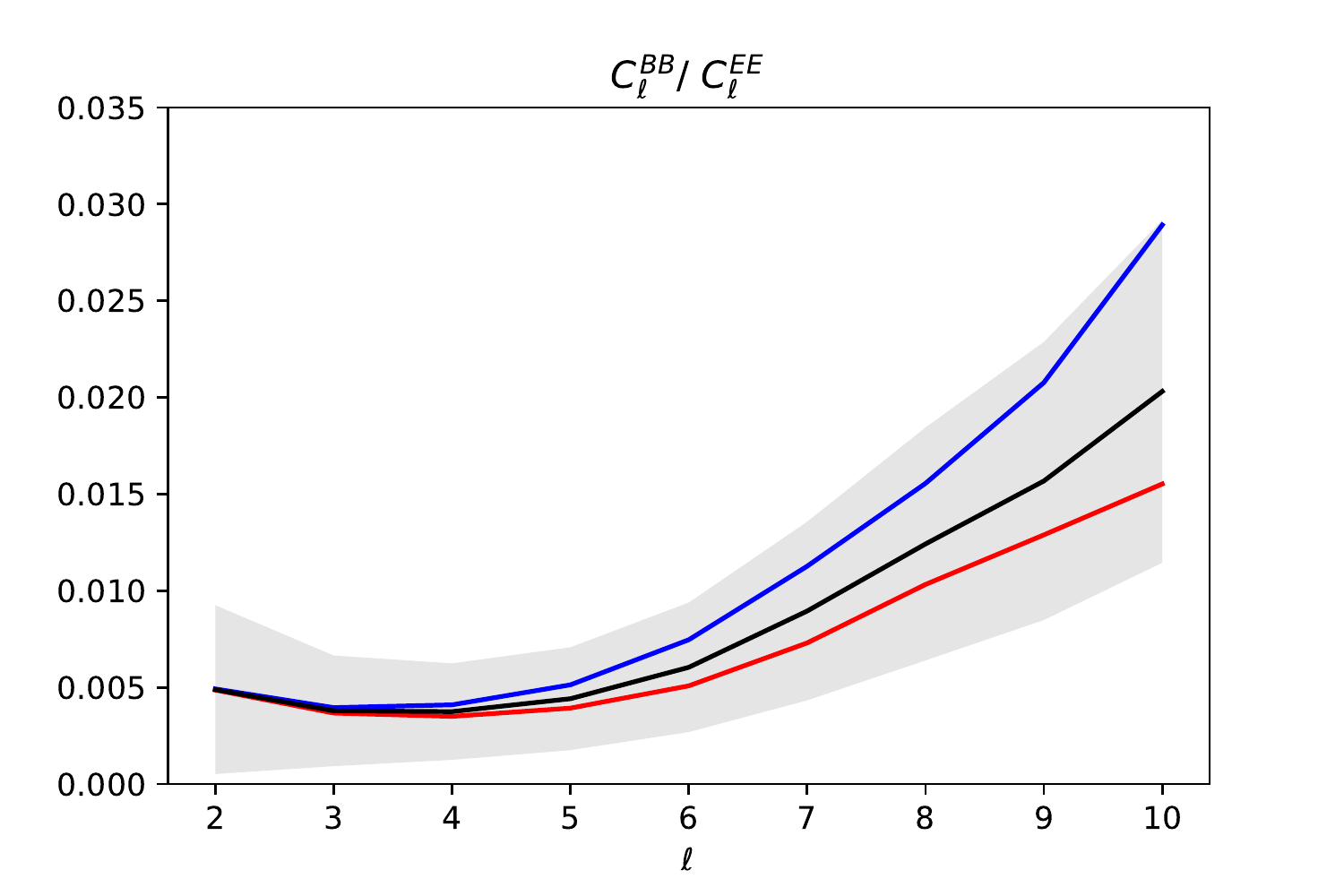}
\quad
\includegraphics[width=50mm]{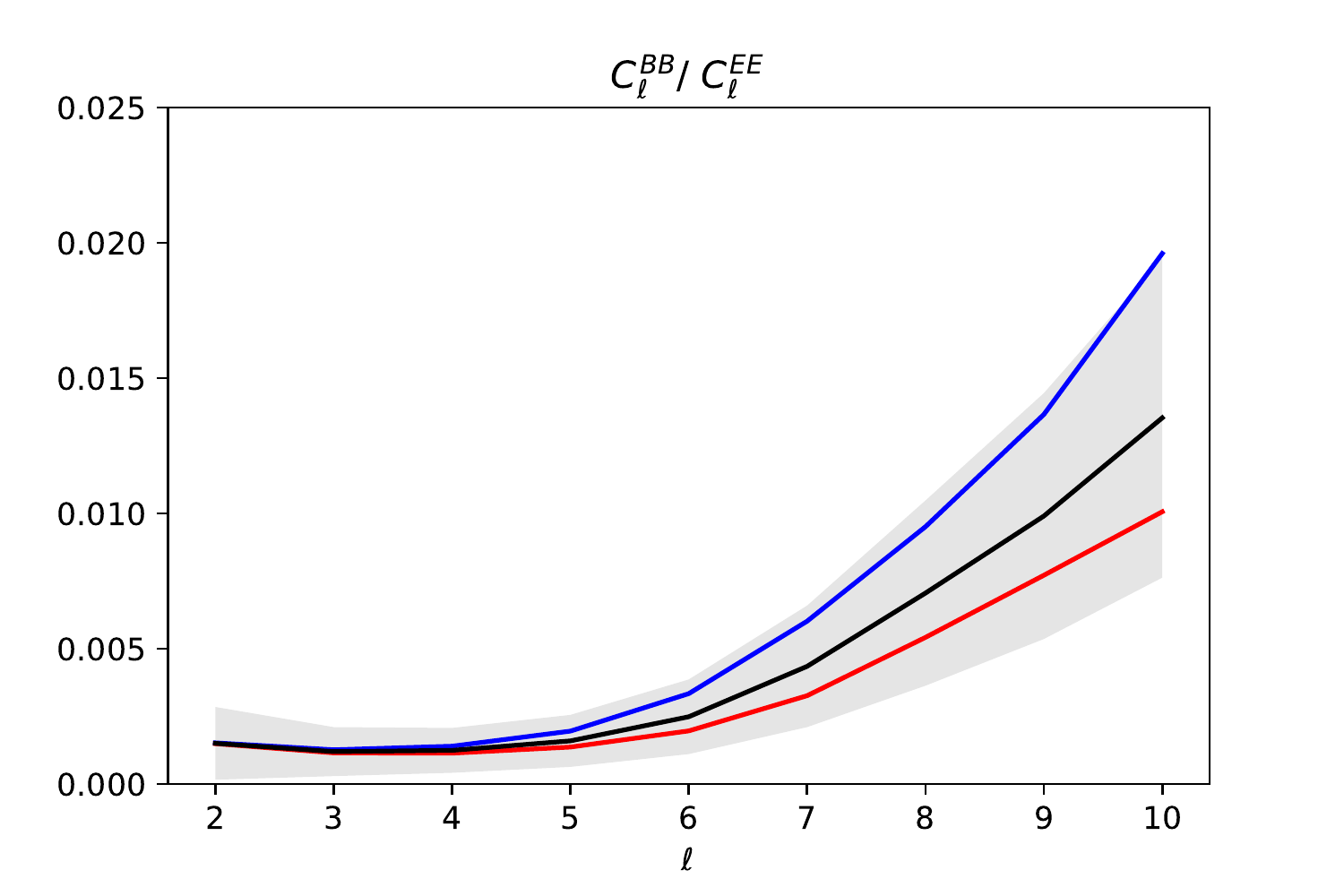}
\caption{
The ratios of $C^{\rm BB}_\ell / C^{\rm EE}_\ell$ changing optical depth $\tau$.
The same color conventions as in the right panel of Fig.~\ref{fig:ionization-functions}.
The value of $r=0.03, 0.01$ and $0.003$ from left to right.
The shaded areas indicate the cosmic variance corresponding to the black line.
}
\label{fig:ratio-changing-tau}
\end{figure}

\begin{figure}[t]
\centering
\includegraphics[width=50mm]{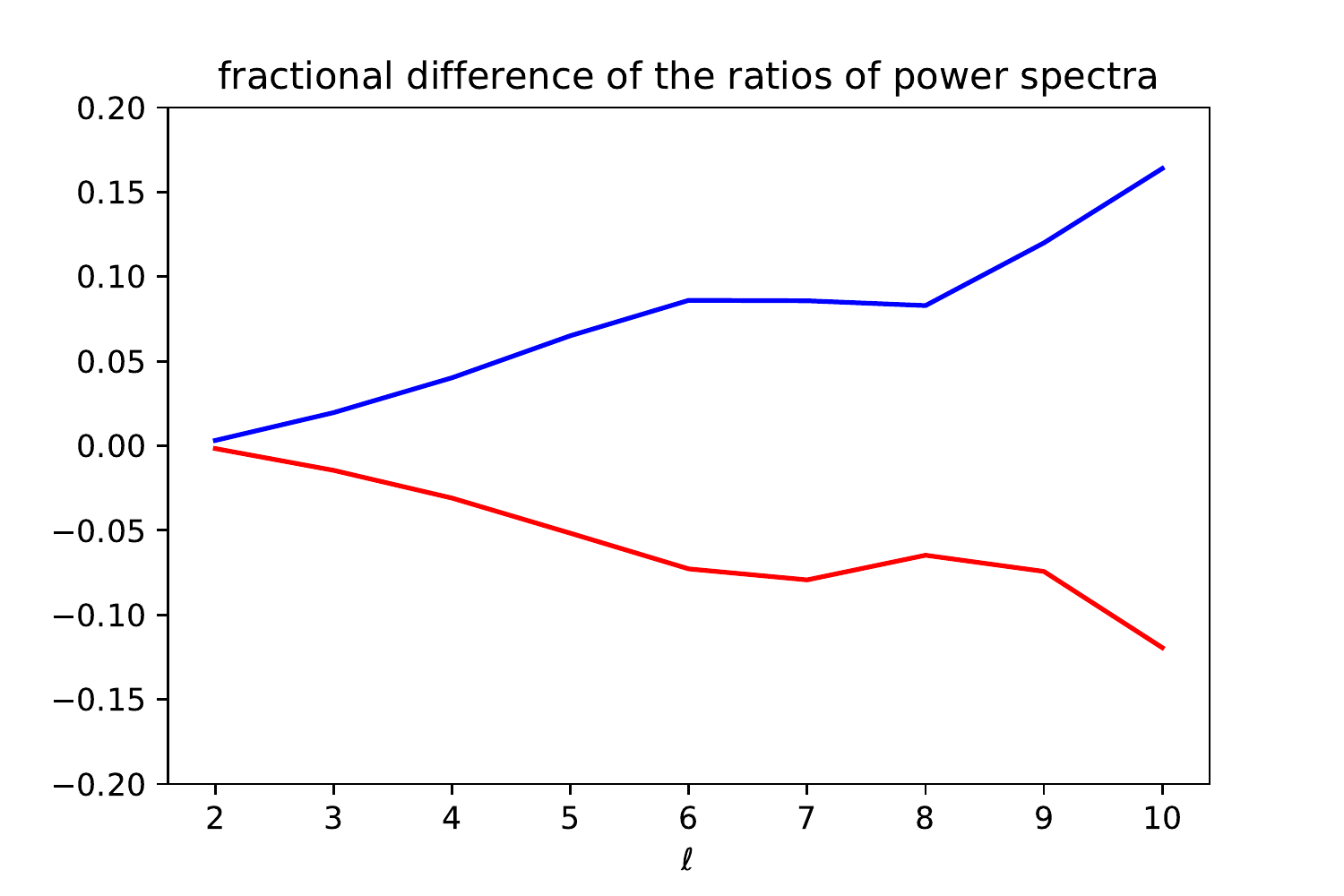}
\quad
\includegraphics[width=50mm]{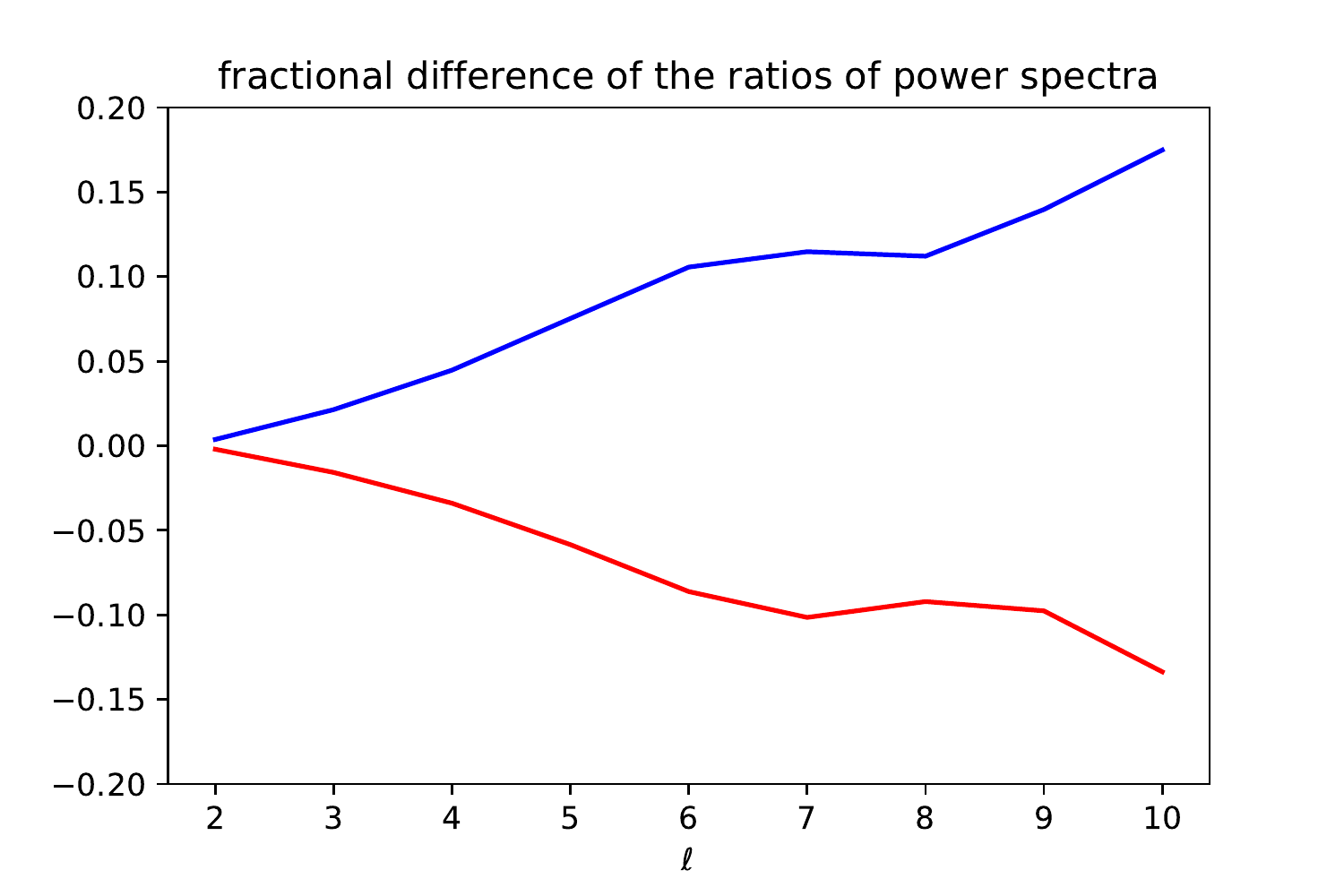}
\quad
\includegraphics[width=50mm]{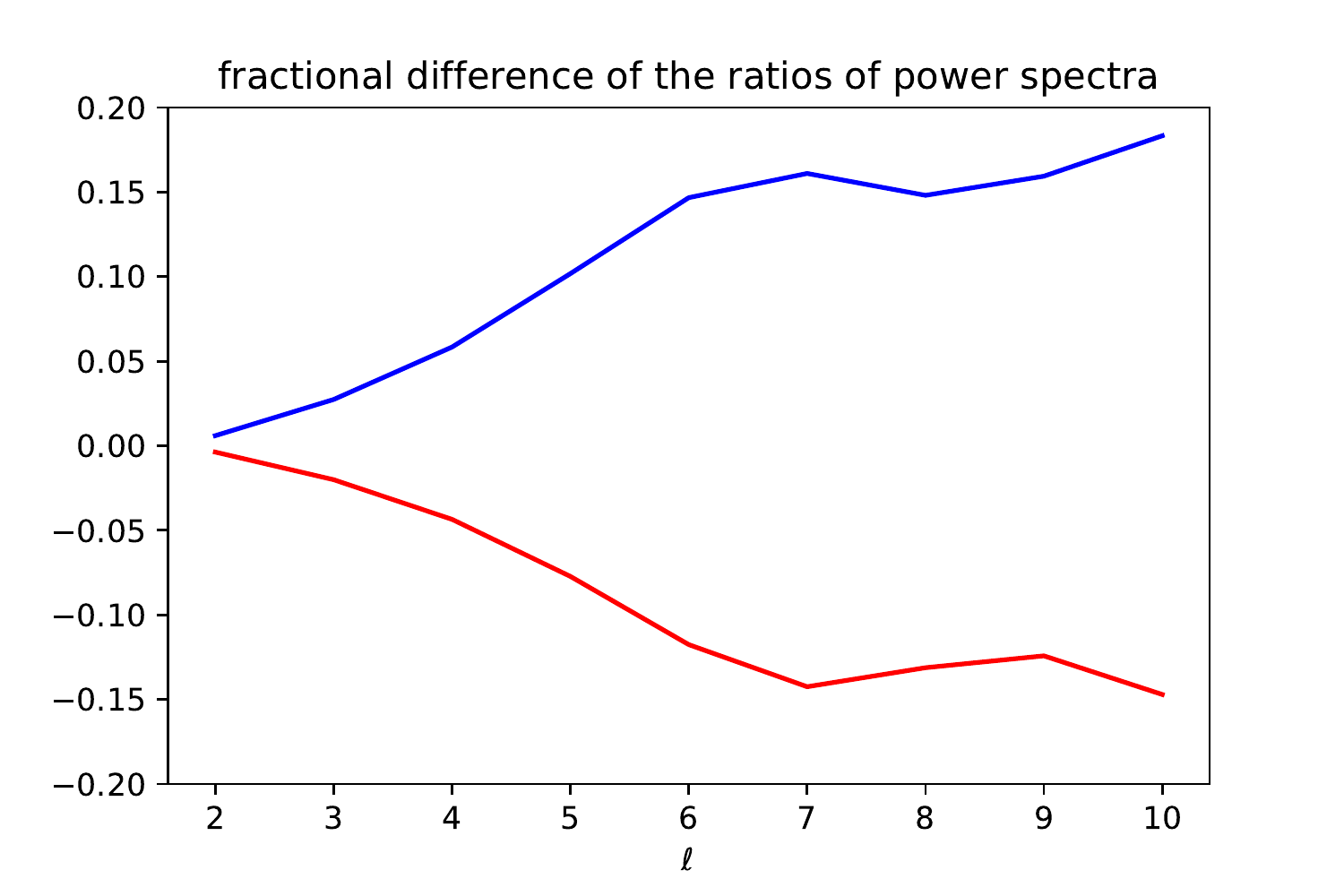}
\caption{
The fractional differences of ratio $C^{\rm BB}_\ell / C^{\rm EE}_\ell$ between
 the case of $\tau = 0.047$ and $\tau = 0.054$ (blue),
 and between
 the case of $\tau = 0.061$ and $\tau = 0.054$ (red).
The value of $r=0.03, 0.01$ and $0.003$ from left to right.
}
\label{fig:diff-ratio-changing-tau}
\end{figure}

We next consider the case of changing optical depth $\tau$ with fixed $\Delta z_{\rm rei}=0.5$.
Figs.~\ref{fig:EE-power-changing-tau} and \ref{fig:BB-power-changing-tau}
 show the EE and BB power spectra.
In general smaller $\tau$ give smaller magnitudes of EE and BB powers in $\ell \lesssim 10$.
We see that
 the $\tau$-dependence is rather large in all values of $r$,
 though it is almost within the cosmic variance.
Fig.~\ref{fig:ratio-changing-tau} shows the ratio $C^{\rm BB}_\ell / C^{\rm EE}_\ell$.
We see that
 the values of the ratio with smaller $\tau$ give larger values than those with larger $\tau$,
 which is the opposite behavior of the individual magnitudes of the EE and BB powers.
This is because that
 the effect of changing $\tau$ to the EE power is larger than that to the BB power
 in all range of $\ell \lesssim 10$.
For smaller values of $\ell$
 the $\tau$-dependence is smaller in all values of $r$.
Fig.~\ref{fig:diff-ratio-changing-tau} shows the fractional differences of the ratios
\begin{equation}
 \frac{\left( \frac{C^{\rm BB}_\ell}{C^{\rm EE}_\ell} \right)_{\tau = 0.047}
       - \left( \frac{C^{\rm BB}_\ell}{C^{\rm EE}_\ell} \right)_{\tau = 0.054}}
      {\left( \frac{C^{\rm BB}_\ell}{C^{\rm EE}_\ell} \right)_{\tau = 0.047}
       + \left( \frac{C^{\rm BB}_\ell}{C^{\rm EE}_\ell} \right)_{\tau = 0.054}}
\quad {\rm and} \quad
 \frac{\left( \frac{C^{\rm BB}_\ell}{C^{\rm EE}_\ell} \right)_{\tau = 0.061}
       - \left( \frac{C^{\rm BB}_\ell}{C^{\rm EE}_\ell} \right)_{\tau = 0.054}}
      {\left( \frac{C^{\rm BB}_\ell}{C^{\rm EE}_\ell} \right)_{\tau = 0.061}
       + \left( \frac{C^{\rm BB}_\ell}{C^{\rm EE}_\ell} \right)_{\tau = 0.054}}
\end{equation}
 with blue and red lines, respectively.
The $\ell$-dependence of fractional differences is larger for smaller value of $r$.
We find that the $\tau$-dependence
 of the value of ratio $C^{\rm BB}_\ell / C^{\rm EE}_\ell$ for $\ell < 5$ is small,
 less than 5\% in fractional difference with fixed $\Delta z_{\rm rei}=0.5$.
The values of the ratio
 $C^{\rm BB}_\ell / C^{\rm EE}_\ell$ with $\ell=2$ divided by tensor-to-scalar ratio $r$
 are $0.49$, $0.49$ and $0.50 \sim 0.51$ for $r=0.03, 0.01$ and $0.003$, respectively,
 in accord with the analytical estimate in the previous section again.
It is remarkable that
 $\tau$-dependence of the values of ratio $C^{\rm BB}_\ell / C^{\rm EE}_\ell$
 becomes smaller for smaller $\ell$,
 though the values of $C^{\rm BB}_\ell$ and $C^{\rm EE}_\ell$ themselves
 largely depend on the values of $\tau$ even for small $\ell$.


We conclude that the reionization process dependence of the ratio $C^{\rm BB}_\ell / C^{\rm EE}_\ell$,
 which is dominated by the ambiguity of the value of optical depth,
 is small for $\ell < 5$ with fractional difference less than 5\%.
The reason is that
 the polarization for small $\ell$ is produced in late time, or at small $z$,
 where the reionization process has been almost finished.
The ionization function $x_e(z)$ at small redshift $z \lesssim 6$ can not be changed so much,
 because it is known that the reionization should be finished at least $z \simeq 6$
 by the observation of Gunn--Peterson trough \cite{Gunn:1965hd,Choudhury:2020vzu,Qin:2021gkn}.
Considering the ratio $C^{\rm BB}_\ell / C^{\rm EE}_\ell$ further reduces the dependence,
 since these power spectra have the same origin, except for larger $\ell$,
 where the EE power spectrum includes non-negligible primordial component
 and the BB power spectrum includes non-negligible component which is generated
 through the gravitational lensing effect from the E-mode component. 
Therefore, it is reasonable
 that the value of the ratio $C^{\rm BB}_\ell / C^{\rm EE}_\ell$ with the smallest $\ell=2$ is stable.
It is interesting that the stable value is the half of tensor-to-scalar ratio,
 though there is a contamination by cosmic variance as seen in
 Figs.~\ref{fig:ratio-changing-slope} and \ref{fig:ratio-changing-tau}.
This fact can be utilized in the determination of tensor-to-scalar ratio with various future observations.

Before closing this section we discuss the impact of a variation in spectral indices.
We may vary the value of $n_s$
 within the range of $n_s = 0.054 \pm 0.007$ by the Planck collaboration \cite{Planck:2018vyg},
 for example.
On the other hand,
 it is reasonable to fix the value of $n_t$ as $n_t \simeq -r/8$
 following the slow-roll consistency condition assuming slow-roll inflation.
Some numerical calculations result that
 the amount of variation of $C^{\rm EE}_\ell$ is at most $1\%$ at low-$\ell$:
 larger/smaller $n_s$ result smaller/larger $C^{\rm EE}_\ell$.
The values of $C^{\rm BB}_\ell$  at low-$\ell$ are almost independent from the variation of $n_s$.
Therefore,
 the values of $C^{\rm BB}_\ell / C^{\rm EE}_\ell$ at low-$\ell$ vary at most $1\%$,
 and the results in this section are not affected by the present ambiguity of spectral indices
 under the reasonable assumption of slow-roll inflation.

\section{Conclusions}
\label{sec:conclusions}

We have investigated the reionization process dependence of the ratio
 of BB and EE angular power spectra $C^{\rm BB}_\ell / C^{\rm EE}_\ell$ for small $\ell \lesssim 10$.
An analytic estimation has been given first
 using the formulation in large wavelength limit for scalar and tensor perturbations,
 which appropriate for small $\ell$.
We have found qualitatively that
 the reionization process dependence of the ratio is smaller for vary small $\ell$,
 and especially in case of $\ell=2$ the value of the ratio is half of the vale of tensor-to-scalar ratio
 almost independently from the reionization process.
Next a numerical analysis has been given by using CAMB code.
A simple model of ionization function, $x_e(z)$, which is implemented in CAMB code as default, has been used,
 since there is no standard reionization model has been established yet in spite of much efforts until now.
An important assumption is
 the neglect of a long tail of ionization function for $z>10$,
 which may be created by Pop-III stars, though it has not been established yet.
The variation of the reionization process
 has been described by the simple model which has two parameters:
 optical depth $\tau$ and $\Delta z_{\rm rei}$
 which describes the duration of hydrogen ionization in redshift.
The values of the ratio $C^{\rm BB}_\ell / C^{\rm EE}_\ell$ have been calculated
 in the model with six combination of these two parameters. 
The $\Delta z_{\rm rei}$ dependence of the ratio $C^{\rm BB}_\ell / C^{\rm EE}_\ell$
 is very small less than 0.5\% in fractional ratio for $\ell < 5$,
 but the $\tau$ dependence is not very small less than 5\% in fractional ratio for $\ell < 5$.
We have concluded that
 the reionization process dependence of the ratio for $\ell < 5$ is small less than 5\% in fractional ratio,
 which is dominated by the ambiguity of the value of optical depth.
Especially in case of $\ell=2$
 the value of the ratio is half of the value of tensor-to-scalar ratio in good precision.
These numerical results confirm the results of the analytic estimation.

Although these remarkable facts,
 the power spectrum at low-$\ell$ is suffered by cosmic variance in general.
The power spectra $C^{\rm BB}_\ell$ and $C^{\rm EE}_\ell$ are suffered by the cosmic variance,
 and taking the ratio $C^{\rm BB}_\ell / C^{\rm EE}_\ell$ does not reduce the difficulty.
Some accidental anisotropy, like the accidental localization of free electrons, could be canceled,
 because the origins of the B-mode and E-mode polarizations are the same for low-$\ell$.
However,
 the possible accidental anisotropies in scalar and tensor perturbations
 are independent and can not be canceled by taking ratio.
In this perspective the fact that
 the value at $\ell=2$ is half of the value of tensor-to-scalar ratio,
 should be considered with rather large error due to cosmic variance
 and it can not be a way to obtain the value of tensor-to-scalar ratio precisely.
In future
 once the knowledge of the reionization process will be precisely established,
 by the observation of 21cm signal, for example,
 including the precise determination of the optical depth, and also
 both EE and BB polarization power spectra at low-$\ell$ will be precisely measured by LiteBIRD, for example,
 the results in this article will be useful for a consistency check
 and contribute to strengthen the understanding of physics behind these phenomena.
Once the value of tensor-to-scalar ratio is determined,
 the value of the ratio at $\ell=2$ can be a reference of the magnitude of cosmic variance.

\section*{Acknowledgments}

This work was supported in part by JSPS KAKENHI Grant Number 19K03851.

\end{document}